\documentclass[12pt,preprint]{aastex}
\usepackage{graphicx}
\usepackage{amssymb}
\usepackage{color}
\usepackage{natbib}
\usepackage{amsmath}
\shorttitle{Radiative Cooling Tests}
\shortauthors{Boley et al.\ }

\begin{document}
\title{3D Radiative Hydrodynamics for Disk Stability Simulations: A Proposed Testing Standard
and New Results.}

\author{Aaron C.\ Boley, Richard H.~Durisen} 
\affil{Indiana University, Astronomy Department, 727 E.\ Third St., Bloomington, IN 47405-7105}
\email{acboley@astro.indiana.edu}

\author{\AA ke Nordlund}
\affil{Neils Bohr Institute, Copenhagen University, Juliane Maries Vej 30, 2100 Copenhagen \O, Denmark}

\author{Jesse Lord} 
\affil{Whitman College, Physics Department, 345 Boyer Avenue, Walla Walla, Washington 99362}

\begin{abstract} Recent three-dimensional radiative hydrodynamics simulations of protoplanetary disks report disparate disk behaviors, and these differences involve the importance of convection to disk cooling, the dependence of disk cooling on metallicity, and the stability of disks against fragmentation and clump formation. To guarantee trustworthy results, a radiative physics algorithm must demonstrate the capability to handle both the high and low optical depth regimes.  We develop a test suite that can be used to demonstrate an algorithm's ability to relax to known analytic flux and temperature distributions, to follow a contracting slab, and to inhibit or permit convection appropriately.  We then show that the radiative algorithm employed by Meji\'a (2004) and Boley et al.~(2006) and the algorithm employed by Cai et al.~(2006) and Cai et al.~(2007, in prep.) pass these tests with reasonable accuracy.  In addition, we discuss a new algorithm that couples flux-limited diffusion with vertical rays, we apply the test suite, and we discuss the results of evolving the Boley et al.~(2006) disk with this new routine.  Although the outcome
is significantly different in detail with the new algorithm, we obtain the same qualitative answers.  Our
disk does not cool fast due to convection, and it is stable to fragmentation.  We find an effective $\alpha\approx 10^{-2}$.  In addition, transport is dominated by low-order modes.
\end{abstract}

\keywords{accretion, accretion disks --- convection --- hydrodynamics --- instabilities --- solar system: formation }

\section{INTRODUCTION}

Despite numerous studies of gravitational instabilities in protoplanetary disks (e.g., Boss 1997, 1998, 2001, 2002, 2005; Pickett et al.~1998, 2000b, 2003; Balbus \& Papaloizou 1999; Nelson et al.~2000; Gammie 2001; Johnson \& Gammie 2003; Rice et al.~2003, 2004, 2005; Mayer et al.~2004, 2007; Lodato \& Rice 2004; Mej\'ia et al.~2005; Rafikov 2005; Durisen et al.~2005; Cai et al.~2006, Boley et al.~2006), researchers still disagree on several key issues (see Durisen et al.~2007 for a review).  Specifically, recent three-dimensional radiative hydrodynamics simulations of protoplanetary disks with gravitational instabilities (GIs) show very different disk evolution, where the differences involve the importance of convection in disks, the dependence of disk cooling on metallicity, and the stability of disks against fragmentation and clump formation (Boss 2004, 2005; Cai et al.~2006; Boley et al.~2006; Mayer et al.~2007).

As we demonstrate herein, disk evolution is sensitive to the details of radiative physics algorithms.  Therefore, before disk evolution can be addressed by any radiative hydrodynamics code, the code's implementation of radiative physics must be compared to analytic test cases so the limitations of the simulation are understood. In this paper we present such test cases, and we challenge all researchers who use radiative hydrodynamics to study disk evolution to run their algorithms through these tests.  We use this test suite to evaluate the accuracy of the radiative routine used by Boley et al.~(2006, hereafter Paper I), who employ the routine developed by Mej\'ia (2005, hereafter the M2004 scheme), and we recompute part of the disk simulation in Paper I with our new radiative scheme (hereafter the BDNL scheme), which uses vertical rays coupled with flux-limited diffusion in the radial and azimuthal directions.

This paper is laid out as follows.  We describe the hydrodynamics code and our new radiative cooling algorithm in \S 2 and 3, respectively.  In \S 4 we present the radiative test suite and apply it to the radiative algorithms used in Paper I and in Cai et al.~(2006) and to the new algorithm.  We compare the BDNL simulation with the Paper I simulation in \S 5, and we discuss the role convection plays in disk cooling in \S 6.  The results are summarized in \S 7.  

\section{HYDRODYNAMICS}

The three-dimensional hydrodynamics code with self-gravity is the same as that used by Pickett et al.~(1995), Mej\'ia et al.~(2005), Cai et al.~(2006), and Paper I, and it is described in detail in Pickett (1995), Pickett et al.~(1998, 2000a), and Mej\'ia (2004).  The code solves the equations of hydrodynamics in conservative form (see Yang 1992) on a uniform cylindrical grid ($r$, 
$\phi$, $z$), and assumes an ideal gas equation of state with a constant ratio of specific heats $\gamma=5/3$.  The potential due to the mass on the grid is found by using a direct Poisson's equation solver (see Tohline 1980), and the potential on the boundary needed for this solver is calculated using spherical harmonics up to $l=m=10$ (see Pickett et al.~1996). Pickett et al.~(2003) found that when a blob was loaded onto the grid, summing up to $l=m=10$ was sufficient to describe the boundary potential and that the error in the solution was dominated by grid resolution.  The code computes the source and flux terms (Norman \& Winkler 1986) separately in 
an explicit, second-order time integration (van Albada et al.~1982; Christodoulou 1989; Yang 1992), where the advective terms are calculated with a van Leer scheme (Norman \& Winkler 1986).  The outermost radial and vertical grid cells are outflow boundaries.  The innermost radial grid cells are also outflow boundaries where the mass flowing through this boundary is added to the central mass.  We impose a point source gravitational field due to the star, but we keep the star fixed at the center.  We choose to hold it fixed because moving the star to a position that explicitly keeps the center of mass at the grid center (Boss 1998) may also cause spurious dynamics.  Improvements to the code that allow for proper star-disk interactions are being developed.

We include the effects of shock heating by artificial bulk viscosity through a second-order Neumann \& Richtmeyer scheme (see Pickett 1995).   This artificial viscosity ensures that the jump conditions are satisfied 
by adding the correct amount of entropy to the gas. For more details on the implemented AV scheme, we refer the reader to Pickett (1995) and to Pickett et al.~(2000a).

\section{RADIATIVE ROUTINES}

\subsection{The M2004 and C2006 Schemes}

The M2004 scheme is described in detail in Paper I, but we summarize the numerical method here.  In their scheme, flux-limited diffusion is used in the $r$, $\phi$, and $z$ directions on the cylindrical grid everywhere that the vertically integrated Rosseland optical depth $\tau > 2/3$, which defines the disk's interior.  For mass at lower optical depths, which defines the disk's atmosphere, the gas is allowed to radiate as much as its emissivity allows, with the Planck mean opacity used instead of the Rosseland mean opacity.  The disk interior and atmosphere are coupled with an Eddington-like boundary condition over one cell.  This boundary condition defines the flux leaving the interior, which can be partly absorbed by the overlaying atmosphere.  Likewise, feedback from the atmosphere is explicitly used when solving for the boundary flux.  However, cell-to-cell radiative coupling is not explicitly modeled in the disk's atmosphere.  This method allows for a self-consistent boundary condition that can evolve with the rest of the disk.  Cai et al.~(2006) improved the stability of the routine, as described below, by extending the interior/atmosphere fit over two cells (hereafter the C2006 scheme).

A problem with the routines employed by Mej\'ia (2004), Cai et al.~(2006), and Paper I is a sudden drop in the temperature profile where $\tau=2/3$.  The drop is due to the omission of complete cell-to-cell coupling in the optically thin regime $(\tau < 2/3)$.  However, as shown in Paper I, the boundary does permit the correct flux through the disk's interior.  Because the flux through the disk is correct, the temperature drop is mainly a dynamic concern inasmuch as it might seed convection (see Paper I).  In order to obtain the correct flux and temperature profiles, a method for calculating fluxes that takes into account the long-range effects of radiative transfer is required. 

\subsection{The BDNL Scheme}

Consider some column in a disk with fixed $r$ and $\phi$.  Take that column out of context, and imagine that it is part of a plane-parallel atmosphere.  In this case, we can easily describe the heating and cooling by radiation with the method of discrete ordinates (see, e.g., Chandrasekhar 1960; Mihalas \& Weibel-Mihalas 1986).  This method uses discrete angles that best sample the solid angle, as determined by Gaussian quadrature. In a plane-parallel atmosphere, a single ray can provide decent accuracy if the cosine of the angle measured downward from the vertical to the ray is $
\mu =1/\sqrt 3$.  We use this approach to approximate radiative transfer in the vertical direction, and we include flux-limited diffusion  (Bodenheimer et al.~1990)  in the $r$ and $\phi$ directions everywhere that $\tau \ge 1/\sqrt 3$. Naturally, this is only a crude approximation when one places the column back into context.  However, we believe it represents the best implementation of radiative physics for simulating protoplanetary disks with three-dimensional hydrodynamics thus far, because it captures the long-range effects of radiative transfer that are excluded in pure flux-limited diffusion routines.  As we demonstrate later, such coupling can affect disk evolution.

Consider now some incoming intensity $I_-$ and some outgoing intensity $I_+$. In the context of the approximation outlined above, the vertical flux at any cell face can be evaluated by computing the outgoing and incoming rays for a given column and by relating them to the flux with
\begin{equation}
F=2\pi\mu\left(I_+-I_-\right).\label{eq1}
\end{equation}
Once we have vertical fluxes at cell faces, we can compute the vertical component of the divergence of the flux at the cell center by differencing fluxes at cell faces.

We compute the outgoing ray by
\begin{equation}
I_+ = I_+(t_d)\exp (-\Delta t) + \int_{t_u}^{t_d} S(t') \exp(t'-t_d)d t'\rm,
\end{equation}
where $\Delta t=t_d-t_u$, $t_d$ is the optical depth at the base of the cell measured {\it along the ray}, $t_u$ is the optical depth at the top of the cell, and $I_+(t_d)$ is the upward intensity at the base of the cell.  Because we are assuming that each column in the disk is part of a plane-parallel atmosphere, the optical depth along the ray can be computed by $t=\tau/\mu$.

Similar to $I_+$, we define the incoming ray solution across one cell as
\begin{equation}
I_- = I_-(t_u)\exp (-\Delta t ) + \int_{t_d}^{t_u} S(t') \exp(t_u-t')d t'\rm,
\end{equation}
where $I_-(t_u)$ is the incoming intensity at the top of the cell.

The 0th approximation for $S(t)$ is that it is constant over the entire cell.  This approximation leads to 
\begin{eqnarray}
I_+=I_+(t_d)\exp\left(-\Delta t\right)+ S_0(1-\exp(-\Delta t))\\
I_-=I_-(t_u)\exp\left(-\Delta t\right)+ S_0(1-\exp(-\Delta t))\rm,
\end{eqnarray}
and $S_0=\sigma T_0^4/\pi$, where $T_0$ is the temperature at the cell center. 

Because the source function typically changes over a cell, additional complexity is necessary.  Consider a source function that may be represented by the quadratic
\begin{equation}
S(t)=c + bt + at^2.
\end{equation}
To find the constants $c$, $b$, and $a$, we Taylor expand the source function about the optical depth defined at the cell center $t_0$:
\begin{equation}
S(t)\approx \bigg\{ S_0-\frac{dS}{d t}\bigg |_{t_0}t_0+\frac{d^2S}{2dt^2}\bigg |_{t_0}t_0^2 \bigg \}
      +  \bigg \{ \frac{dS}{dt}\bigg |_{t_0}-\frac{d^2S}{dt^2}\bigg |_{t_0}t_0 \bigg \}t
      +  \bigg \{\frac{d^2S}{2dt^2}\bigg |_{t_0} \bigg \}t^2.
\end{equation}
The first term in curly brackets in equation (7) is $c$, the second is $b$, and the third is $a$.  Using equation  (7), we can find solutions for equations (2) and (3) across any given cell (see also Heinemann et al.~2006).
%
%
%
%
However, in order to use equation (7), the source function's derivatives must be evaluated:
\begin{eqnarray}
\frac{dS}{dt}\bigg |_{t_0}=S'_0&=&2\mu \sigma T_0^3\frac{(T_{-1}-T_{+1})}{\pi\rho_0\kappa_0\Delta z}{\rm,~and}\\
\frac{dS'}{dt}\bigg |_{t_0}&=&\mu\frac{(S'_{-1}-S'_{+1})}{2\rho_0\kappa_0\Delta z}.
\end{eqnarray}
Here, the 0 denotes the {\it center} of the cell of interest, the -1 denotes the cell center below the cell of interest, and the +1 denotes the cell center above the cell of interest. This difference scheme is used unless the following conditions are met: (A) If the +1 cell's density is below the cutoff value, i.e., the minimum density at which we still compute radiative physics, or the -1 cell's density is below the cutoff value, the derivatives are set to zero, which reduces the solutions for $I_+$ and $I_-$ to equations (4) and (5).  (B) If cell 0 is the midplane cell, i.e., the first cell in the upper plane, a five-point center derivative is used for $S'$, i.e., 
\begin{equation}
\frac{dS}{dt}\bigg |_{t_0}=\mu\sigma T_0^3\frac{8T_0-7T_{+1}-T_{+2}}{3\pi\rho_0\kappa_0\Delta z}\rm,
\end{equation}
unless exception (A) is met.  The simple form of equation (10) is due to the reflection symmetry about the midplane that is built into the grid, which means that the -1 cell's values are equal to the midplane cell's values and that the -2 cell's values are equal to the +1 cell's values.  In addition, the second derivative of the source function at the midplane is taken to be the average of the three-point centered difference method and a forward difference method, i.e., equation (9) is used as one would normally use it to compute the second derivative but that answer is averaged with the derivative obtained by differencing $S'_0$ and $S'_{+1}$.  Various differencing schemes have been tested, and this differencing scheme yields the best results for the widest range of optical depths and cell resolution.

Now that we have a solution for the source function integral, the incoming and outgoing intensities can be computed.  The incoming ray is computed first by summing the solutions to the source function integral as one moves down into the disk along the ray with the previous sum serving as $I_-(t_u)$.  If desired, an incident intensity at $t=0$, as in Cai et al.~(2006), can be added to the solution by extincting the intensity according to the optical depth.  Because reflection symmetry is assumed about the midplane, the incoming intensity solution at the midplane serves as the $I_+(t_d)$ for the outgoing intensity at the midplane.  

For the $r$ and $\phi$ directions, the flux-limited diffusion scheme described by Paper I is employed when the following conditions are met: (A) The vertical Rosseland mean optical depth at the center of the cell of interest is greater than or equal to $1/\sqrt 3$.  This condition ensures that we only compute flux-limited diffusion where photons moving vertically have less than about a 50\% chance of escaping.  (B) The cells neighboring the cell of interest  also have a $\tau \ge 1/\sqrt3$.  This should ensure that the code only calculates temperature gradients between relevant cells; the flux at this cell face is accounted for in the total energy loss (gain) of the system.  If a neighboring cell has a $\tau < 1/\sqrt3$, then the flux at that face is taken to be the vertical flux through the first cell that is below $\tau < 1/\sqrt3$ in the column of interest.  These conditions are similar to those employed by Mej\'ia (2004), Cai et al.~(2006), and Paper I.  Once fluxes have been determined for all cell faces, the divergence of the flux can be calculated with
\begin{equation}\nabla\cdot {\bf F}=\frac{\partial \left(r F_r\right)}{r\partial r} + \frac{\partial F_{\phi}}{r \partial \phi} +
                              \frac{\partial F_z}{\partial z}.\end{equation}

Because the radiative time scale can be much shorter than the hydrodynamic time scale, we employ a radiative cooling (heating) limiter with magnitude
\begin{equation}\nabla\cdot{\bf F}_{\rm limiter} = \frac{\epsilon}{ 0.1\ {\rm orp}},\end{equation}
where 1 orp $\approx 253$ yr is the initial outer rotation period of the disk near 33 AU.   In a similar fashion, we define some maximum heating due to artificial bulk viscosity
\begin{equation}\Gamma_{\rm limiter} = \frac{\epsilon}{ 0.1\ {\rm orp}}\end{equation}
because numerical instabilities can arise without this and/or the time step can drastically decrease due to extremely high temperatures in typically uninteresting parts of the calculation. We monitor the number of cells affected by these limiters during the calculation, and during the asymptotic phase, less than a few percent of the relevant AV heated cells are limited and less than a percent of the relevant radiatively cooling cells are limited.

Finally, we note that we only use Rosseland optical depths, with the opacity evaluated at the cell's local temperature. This is a step backwards from the M2004 scheme, which employs Planck means for regions where the Rosseland $\tau\lesssim 2/3$.  Simply switching opacities for different regions of the disk can lead to erroneous physics when tracing rays in our scheme, e.g., changing the location of the photosphere of the disk.  Regardless, as demonstrated in \S 4, the BDNL scheme performs better overall.  A method that can smoothly transition between the two mean opacities, such as some weighted average based on the midplane Rosseland optical depth, would improve treatment of the opacities by the BDNL  scheme, but has not been attempted here.

\section{RADIATIVE TESTS}

An analytic solution to a relevant test problem must be found in order to evaluate the accuracy of radiative transport algorithms in disks.  In this section, we propose a toy problem based on Hubeny (1990) that can be used to test the accuracy of a radiative routine. 

Consider a plane-parallel slab with constant vertical gravitational acceleration $g$ but with a midplane about which reflection symmetry is assumed.  This situation is not meant to be realistic, but that is okay for the following tests.  Assume there is some heating mechanism that produces a known distribution of astrophysical flux once the system reaches hydrostatic and radiative equilibrium; in equilibrium, energy transport is only vertical. Make the ansatz that the vertical astrophysical flux has the form 
\begin{equation} F_z \left( \tau \right) = F_0\left( 1-\frac{\tau}{\tau_m}\right), \label{eq1}\end{equation}
where $F_0=\sigma T_e^4/\pi$ is the effective astrophysical flux from the atmosphere with effective temperature $T_e$, $\tau$ is the Rosseland mean optical depth measured vertically downward, and $\tau_m$ is the optical depth at the midplane.  This function ensures that the flux goes to zero at the midplane and that $F_z=F_0$ at  $\tau=0$.  The heating term required to achieve this flux distribution is then
\begin{equation} \Gamma = -\pi \frac{\partial F\left(\tau\right)}{\partial z}
                           =\pi F_0 \frac{\rho\kappa}{\tau_m}, \label{eq2}\end{equation}
where $\rho$ is the density at the point of interest and $\kappa$ is the Rosseland mean mass absorption coefficient.

If $\tau_m \gtrsim 10$, the temperature structure may be derived from the flux by using the standard Eddington approximation, which relates the mean intensity $J$ to the astrophysical flux by
\begin{equation}\frac{4}{3}\frac{d J\left(\tau\right)}{d\tau} = F_z\left(\tau\right). \label{eq3}\end{equation}
Integrating equation (16) yields
\begin{equation} T^4=\frac{3}{4}T_e^4\left(\tau\left[1-\frac{\tau}{2\tau_m}\right] + q\right).\label{eq4}
\end{equation}
The constant $q$ can be determined by considering the low optical depth limit.  In that limit, the atmosphere reduces to a standard stellar atmosphere; therefore $q=1/\sqrt3$ (Mihalas \& Weibel-Mihalas 1986).

In the limit that $\tau_m \lesssim 0.5$, the atmosphere approaches an isothermal structure.  Because the source function becomes constant, the observed flux can be found from $F=2\pi \mu I_+$, which gives us the temperature of the isothermal atmosphere:
\begin{equation} T^4=\frac{T_e^4}{2\mu \left(1-\exp\left[-2\tau_m/\mu\right]\right)}\rm ,\end{equation}
where the factor of 2 in the exponential accounts for both sides of the atmosphere, and we explicitly include $\mu$ because we are using the vertically integrated $\tau$.

An additional assumption about the form of the opacity law permits analytic evaluation of the hydrostatic structure of the disk and allows us to control whether the atmosphere will be convective.  For example, we assume that $\kappa$ is constant throughout the disk for two of the tests presented below.  This assumption should make the disk convectively stable (Lin \& Papaloizou 1980; Ruden \& Pollack 1991).  It also makes the relation between pressure and optical depth simple:
\begin{equation}p=\frac{g}{\kappa}\tau. \label{eq5}\end{equation}
Moreover, the midplane pressure is related to the full disk surface density $\Sigma$ by
\begin{equation}p_m=g\frac{\Sigma}{2}.\end{equation}

\subsection{Relaxation Test}

For the relaxation test, we heat the gas slab in accordance with equation (15) to test whether the radiative routine in question allows it to relax to the correct hydrostatic and radiative equilibrium configuration.  By selecting different midplane optical depths, the effects of resolution on the routine can be tested. 

Here, we test the radiative routines used by  M2004, C2006, and BDNL.   The initial condition for the atmosphere is a constant density structure with a $\tau_m = 0.05$, 0.5,  5, 10, and 100.  The target effective temperature of the atmosphere, which controls the magnitude of the heating term, is set to $T_e=100$ K, and $\kappa=\kappa_0=\rm~constant$.  For the discussion that follows, we take the high optical depth regime to indicate where $\tau\ge 1/\sqrt 3$ and the low optical depth regime to indicate where $\tau < 1/\sqrt 3$.   Note that this boundary is slightly different from the M2004 and C2006 schemes; these schemes use $\tau = 2/3 $ to set the boundary between the low and high optical depth regimes, as is done in the standard Eddington solution.

Figure 1 compares the M2004, C2006, and BDNL solutions to the analytic temperature profile and to the flux profile for $\tau_m=100$.  The BDNL routine matches the analytic curves very well. The M2004 routine does well in matching the temperature curve for most of the high optical depth regime, which is resolved by 12 cells, but the low optical depth regime has a sudden temperature drop.  As reported by Paper I, this temperature drop is an artifact of the lack of complete cell-to-cell coupling in the optically thin region.  Despite the temperature drop, the boundary condition between the two regimes typically yields the correct boundary flux, i.e., the flux leaving the high optical depth regime.  An unfortunate result of this boundary condition is that it produces oscillations in the flux profile and in the temperature profile when the low/high optical depth boundary is between two cells or when the entire disk height is only resolved by a few vertical cells.  These numerical oscillations produce artificial heating, and we believe that this behavior contributes to keeping the inner 7 AU of the disk presented in Paper I hot.   Although this is problematic, even when oscillations occur, the time-averaged cooling time is within 10\% of the expected value for the test shown in Figure 1.  The C2006 improvement lessens this problem. 

\begin{figure}[ht]
   \centering
   \includegraphics[width=7cm]{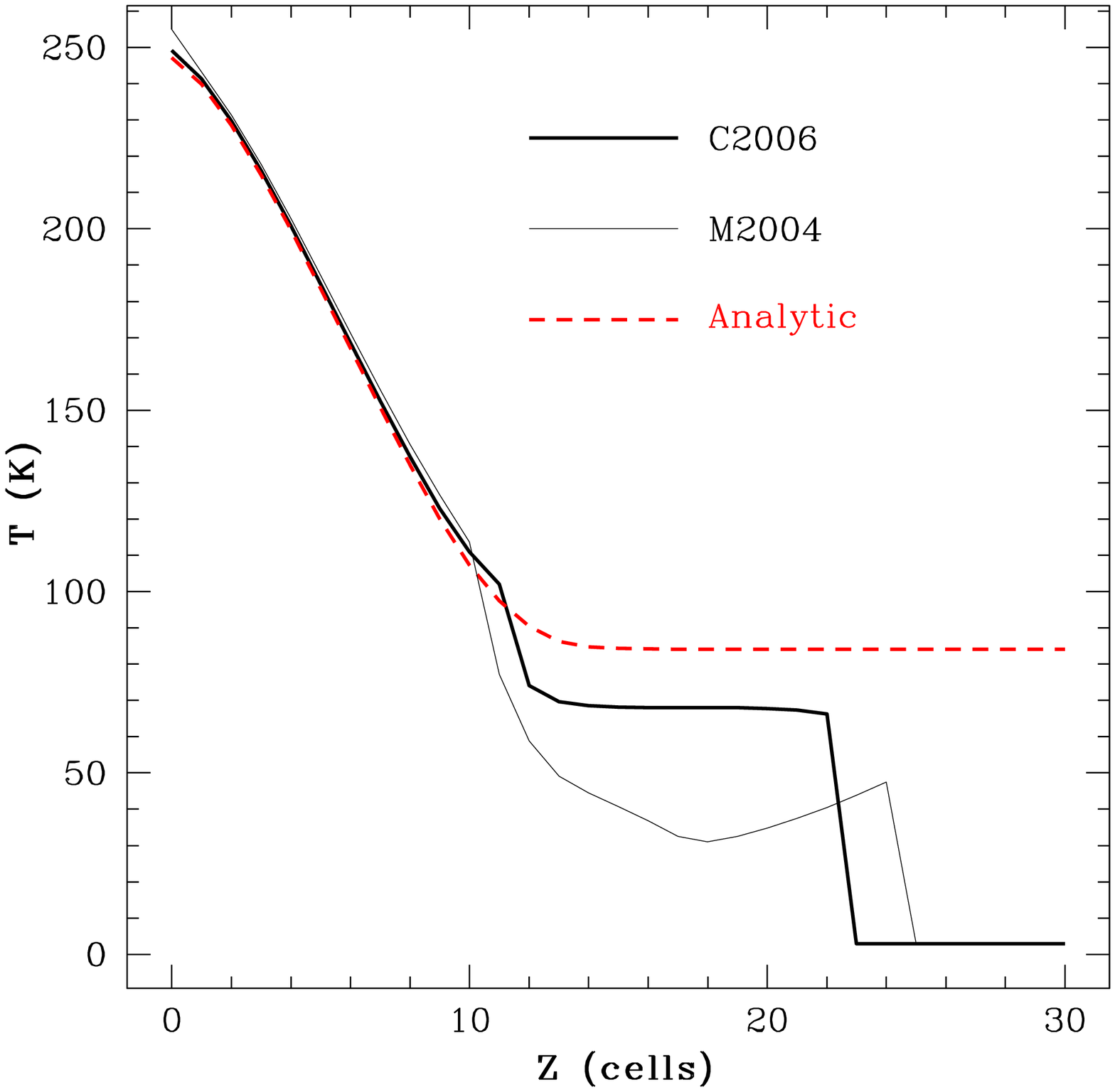}\includegraphics[width=7cm]{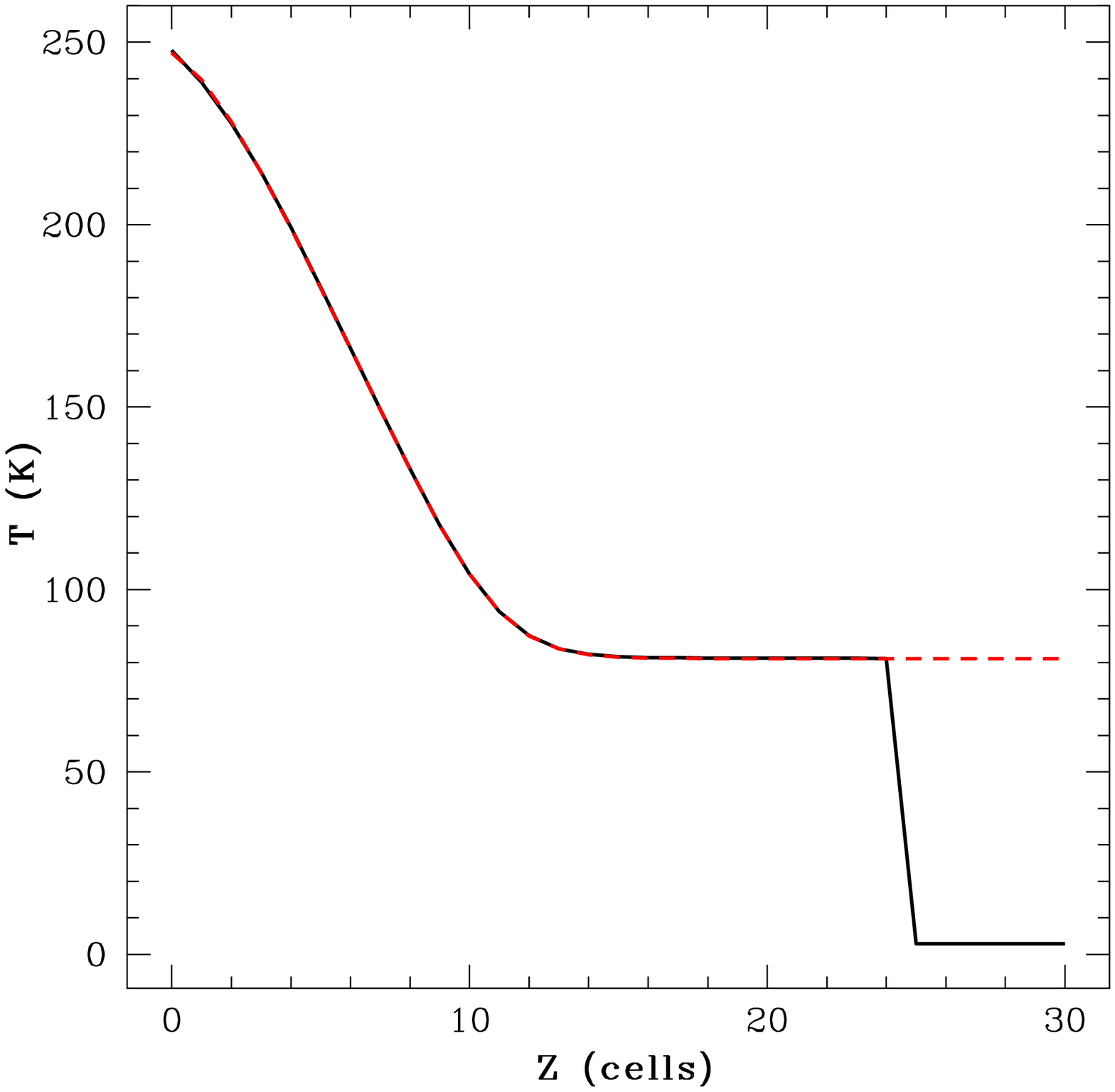} 
   
      \includegraphics[width=7cm] {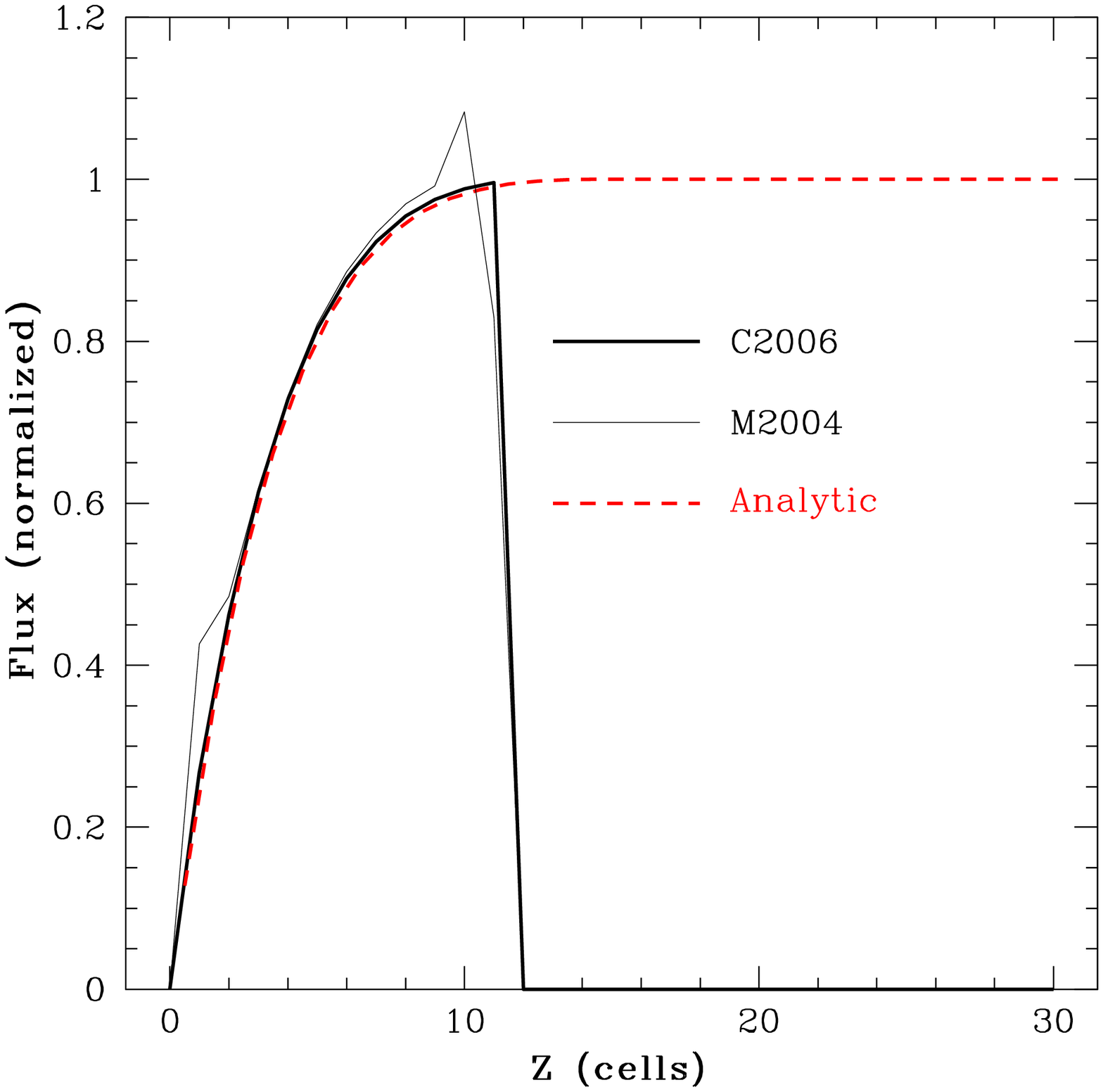}\includegraphics[width=7cm]{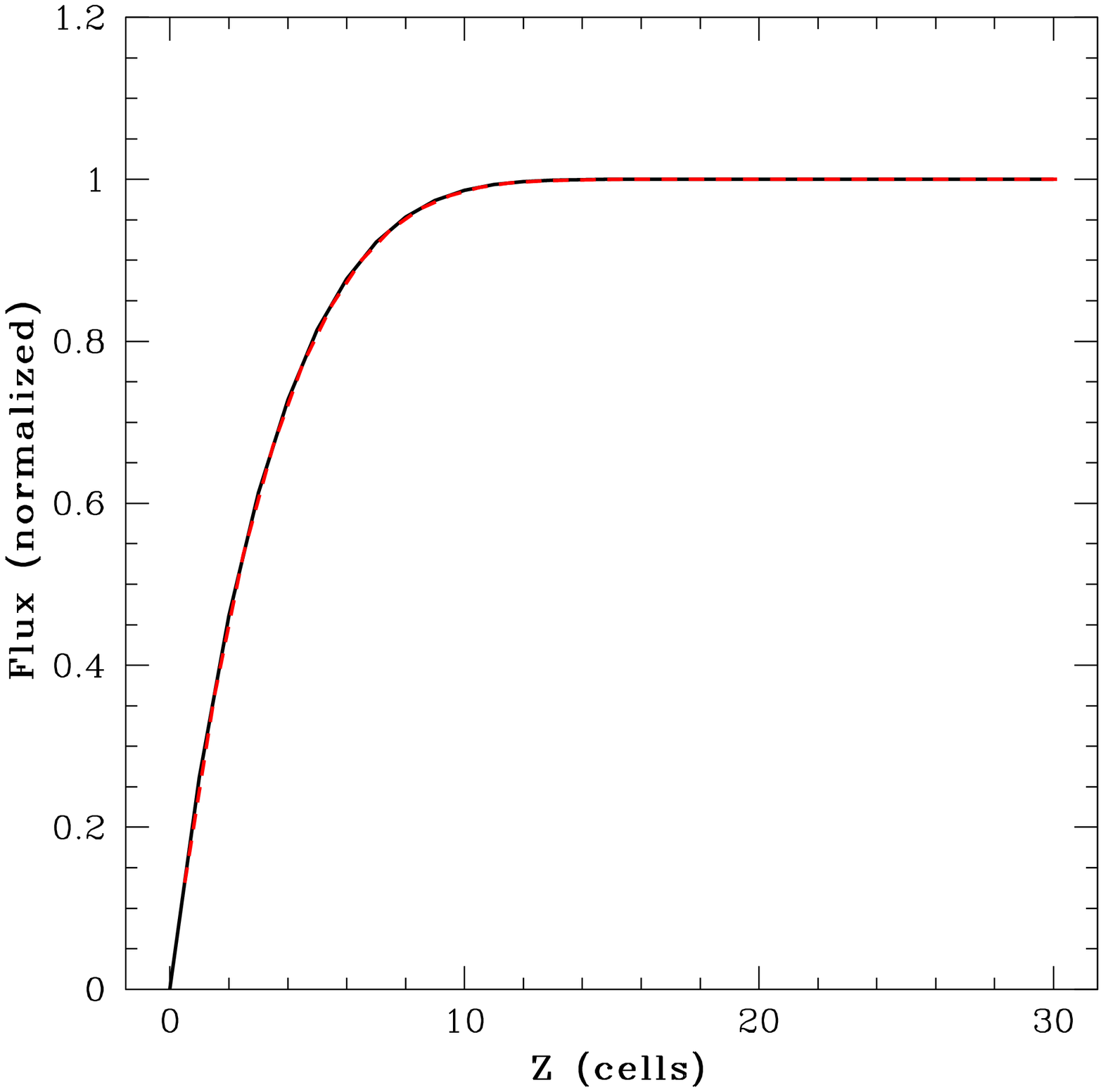}
   \caption{Results of the relaxation test for the M2004, C2006, and BDNL schemes with $\tau_m=100$. Top: The left panel shows the relaxed temperature profile for the M2004 and C2006 routines, while the right panel shows the relaxed temperature profile for the BDNL routine.  In both panels, the analytic curve is represented by the dashed curve.  The first sudden drop in the M2004 and C2006 schemes is due to the lack of complete cell-to-cell coupling required by radiative transfer.  The second drop, which is also in the BDNL profile, is where the density drops to background and where we stop following radiative physics.  Bottom: Similar to the top panels but for the flux profile.  The undulations in M2004's flux profile are believed to be due to the low/high optical depth boundary lying near a cell face intersection.  The C2006 modification helps avoid this problem. Finally, the sudden drop in the M2004 and C2006 flux profile occurs because these schemes only explicitly track the flux through the optically thick disk.   The right panels are the same as Figure 1 in Cai et al.~(2007, in prep.).}
   \label{fig1}
\end{figure}

Figure 2 compares the results of the M2004 and BDNL routines with each other for $\tau_m=10$.  The high optical depth regime is now resolved by only six cells.  Both methods compute the correct flux through the slab. The temperature profiles are both skewed more than in the $\tau_m=100$ case  mainly because the solution deviates from the Eddington approximation, which is used to derive the analytic temperature profile, as $\tau_m$ becomes small.   Figure 3 shows the same comparison as done in Figures 1, 2, and 3 but for $\tau_m=5$.  The high optical depth regime is now resolved by only four cells.  Again, both methods allow for the correct flux through the slab, and the BDNL temperature distribution is close to the analytic value, with slight departures again due mainly to the inaccuracy of the analytic curve.

\begin{figure}[ht]
   \centering
   \includegraphics[width=7cm]{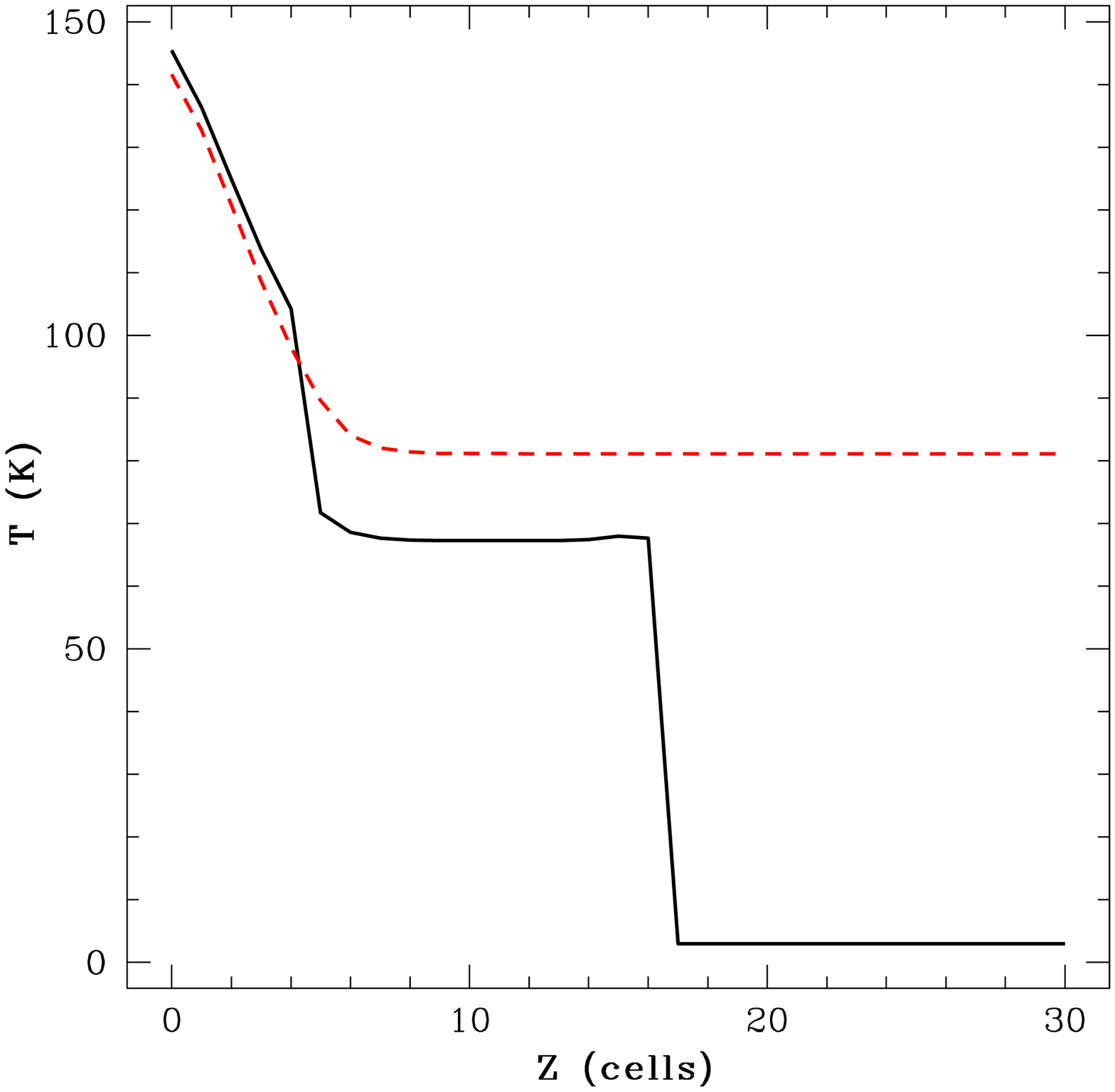}\includegraphics[width=7cm]{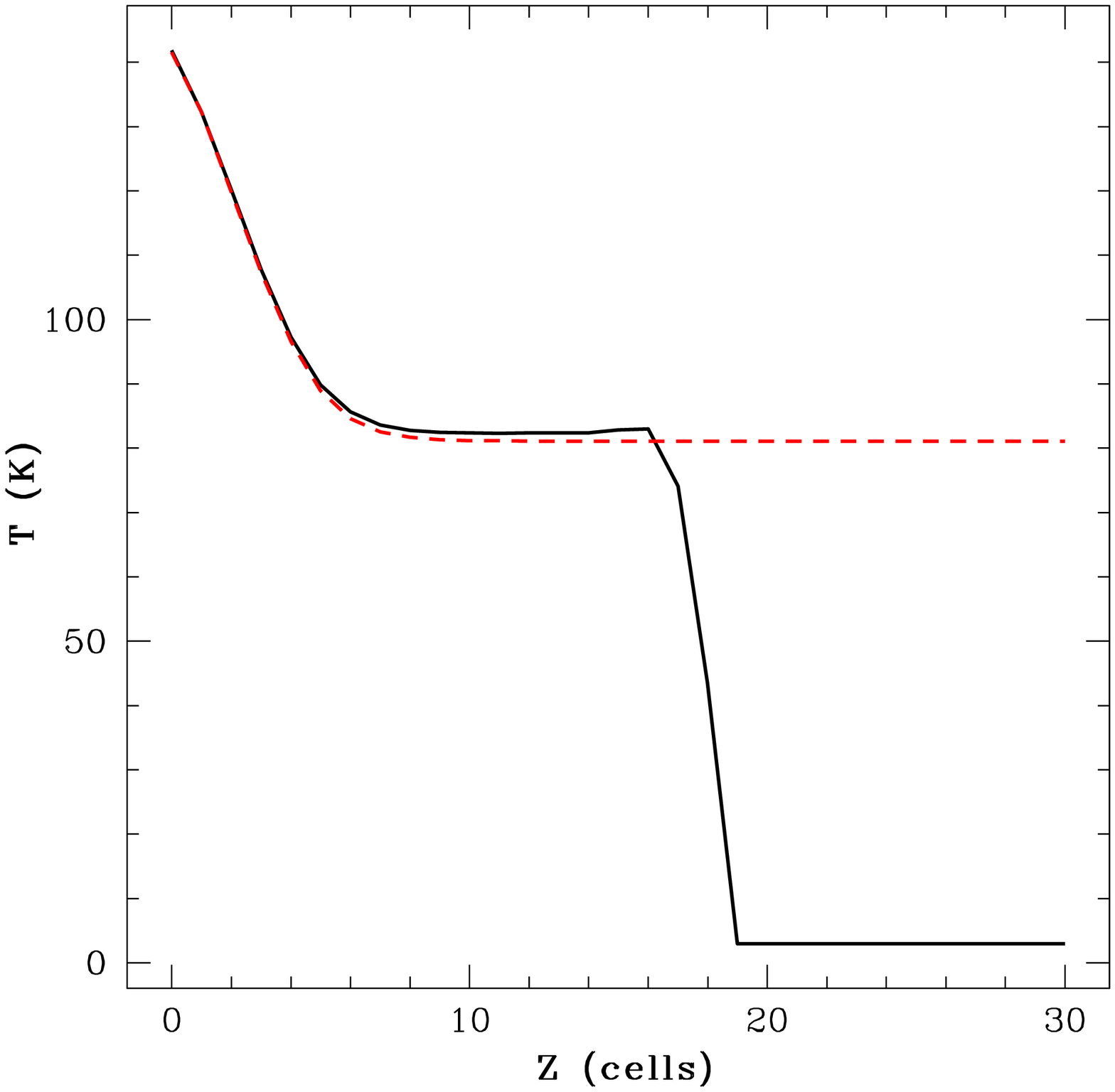} 
   
      \includegraphics[width=7cm] {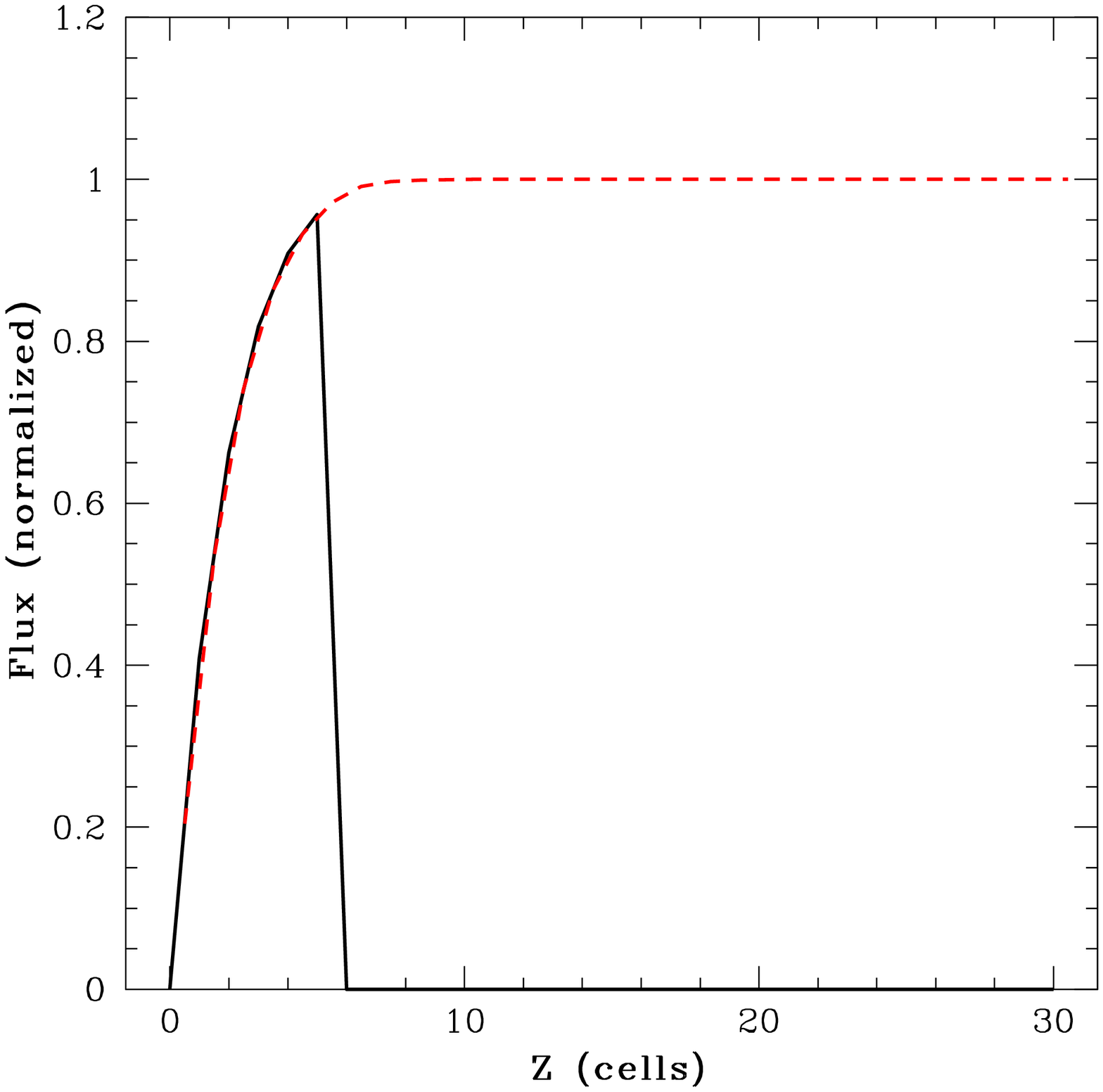}\includegraphics[width=7cm]{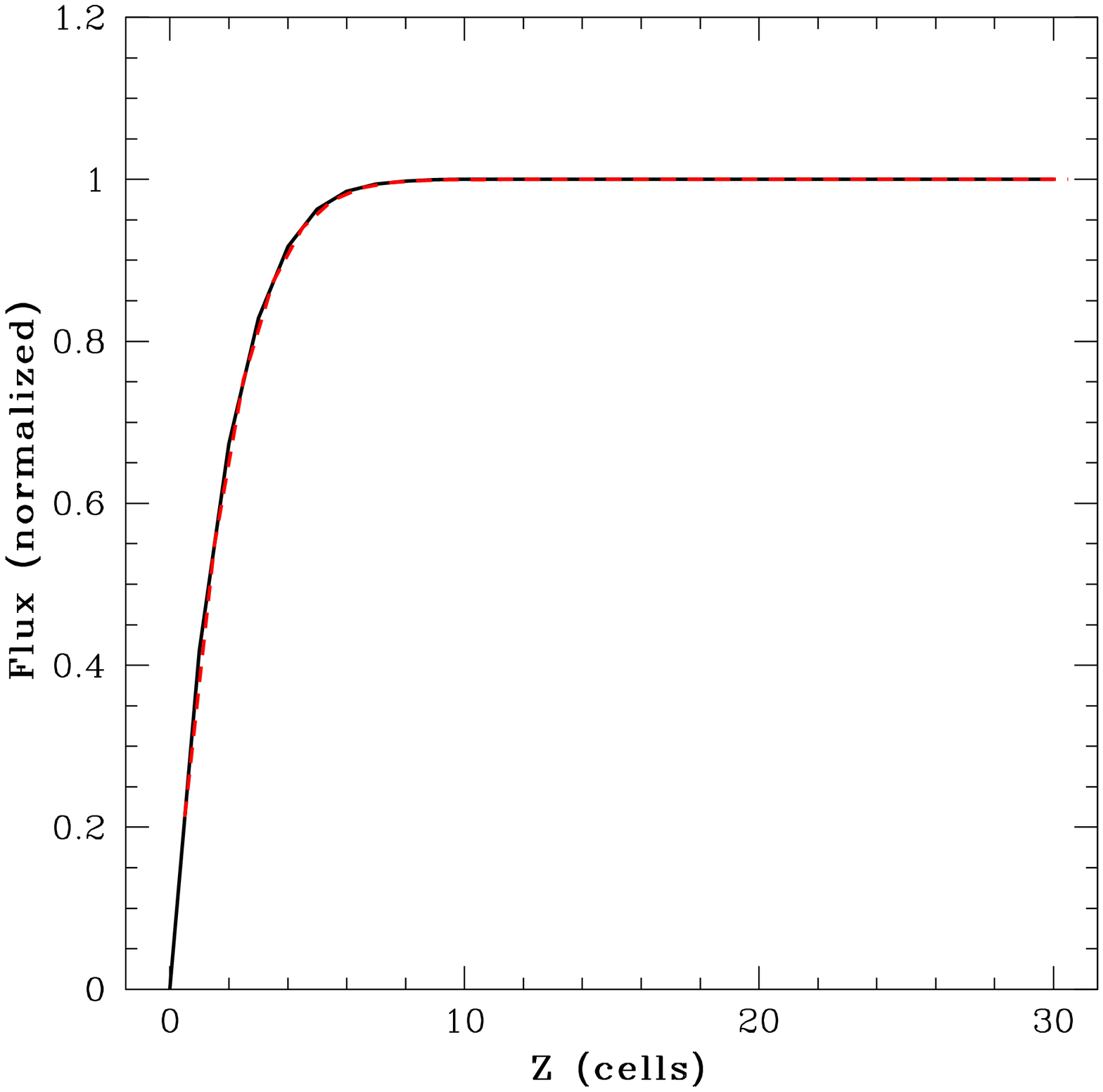}
   \caption{Same as Figure 1, but with $\tau_m=10$ and only for the M2004 and BDNL schemes.  The undulations in the M2004 flux profile are no longer present. The slight departure of the BDNL solution from the analytic temperature curve is mainly due to the breakdown in the Eddington approximation, which is assumed in the analytic temperature profile, as $\tau_m$ becomes small. }
   \label{fig2}
\end{figure}

\begin{figure}[ht]
   \centering
   \includegraphics[width=7cm]{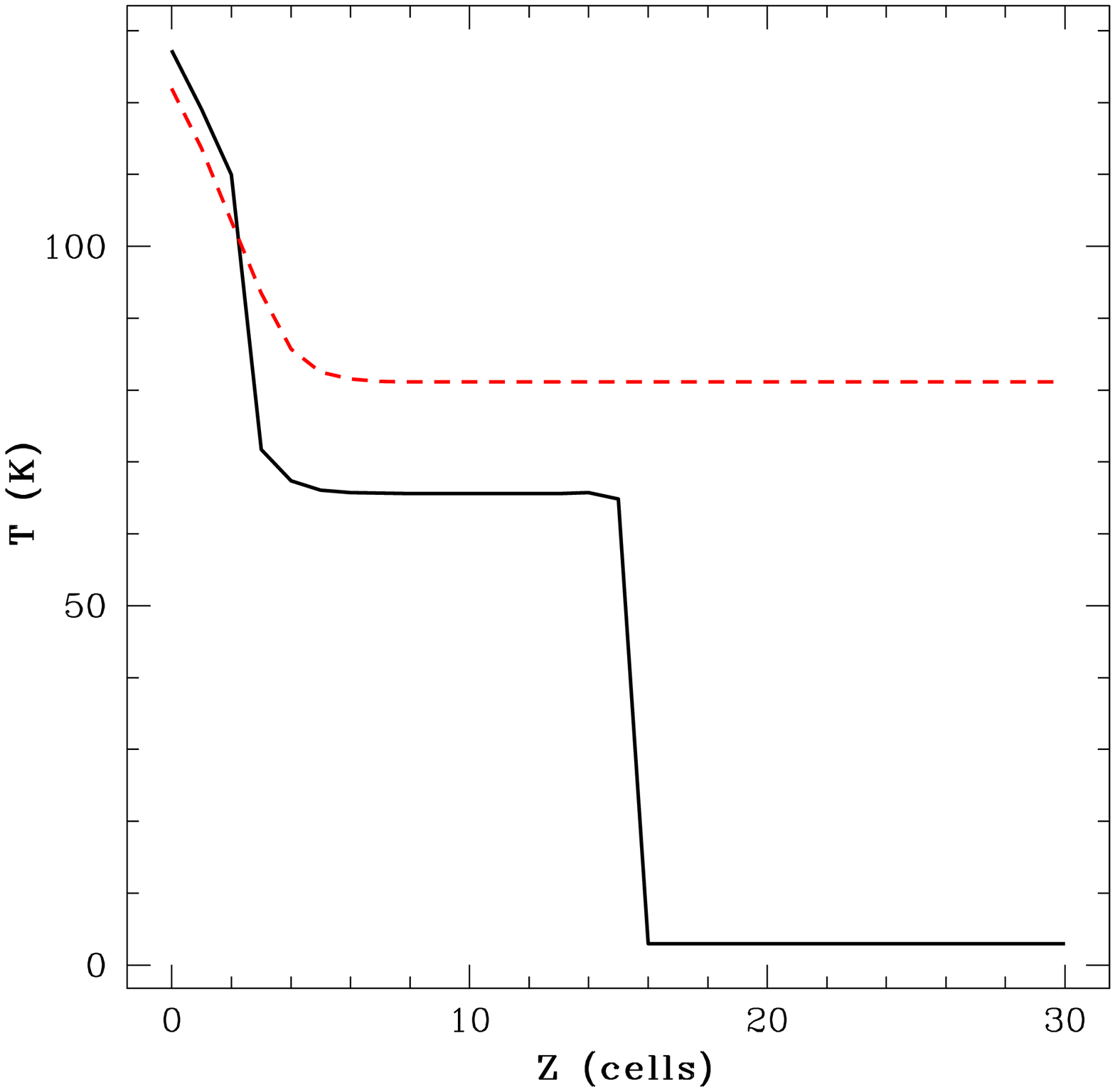}\includegraphics[width=7cm]{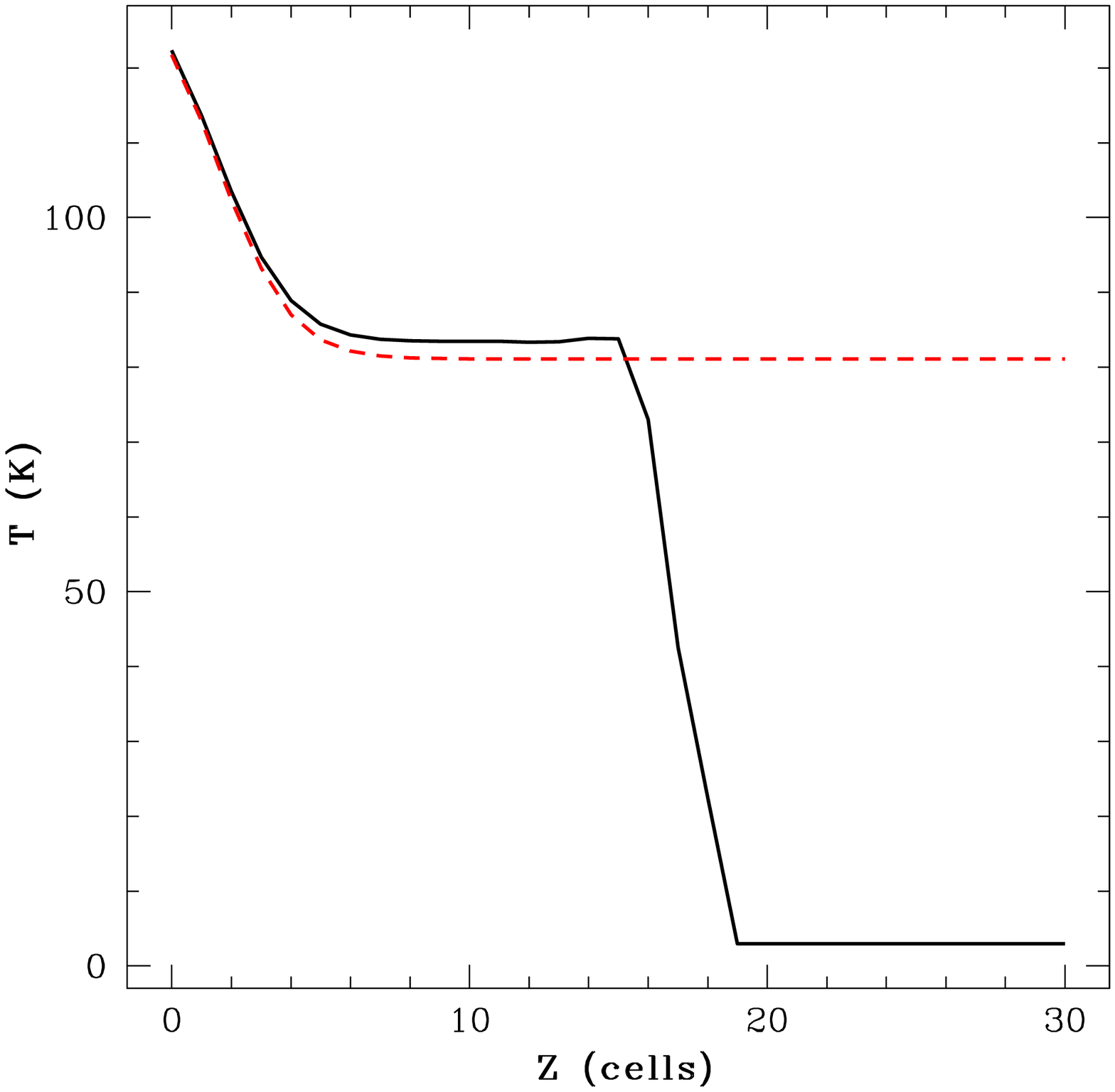} 
   
      \includegraphics[width=7cm] {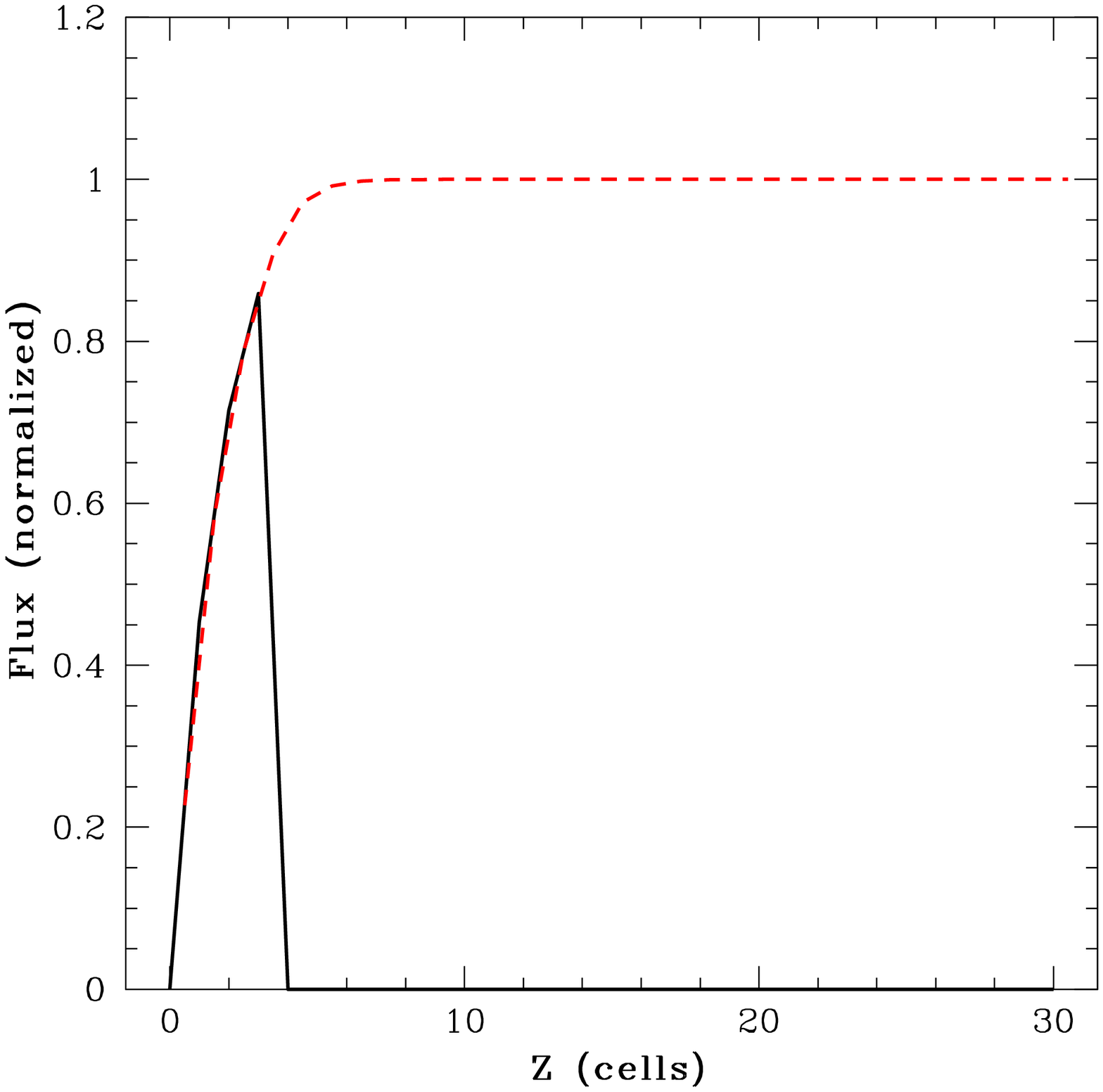}\includegraphics[width=7cm]{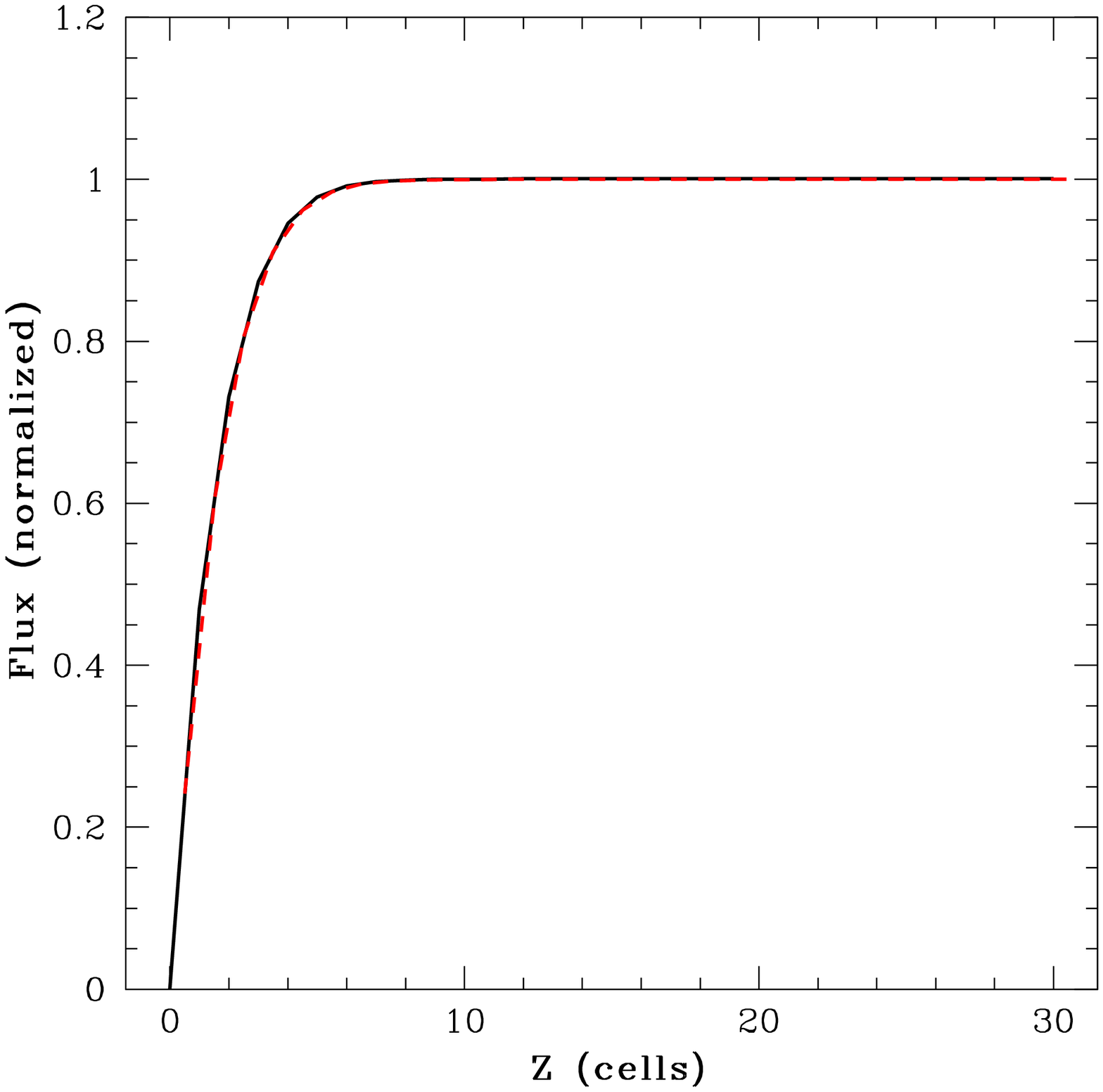}
   \caption{The same as Figure 2, but with $\tau_m=5$. Once again, the depature of the BDNL temperature solution is mainly due to the breakdown of the approximations used to calculate the analytic curve.}
   \label{fig3}
\end{figure}

To demonstrate the accuracy of each algorithm in the low optical depth limit, we show the temperature profiles in Figure 4 for the M2004 routine and the BDNL routine for $\tau_m=0.05$ and 0.5, and we compare each curve with the temperature estimate calculated from equation (18).  The M2004 routine yields temperature profiles that are colder than the expected temperature profiles, and the departure is more severe as $\tau_m$ increases.  This is a result of the lack of complete cell-to-cell coupling in the atmosphere.  The BDNL routine is in excellent agreement with the predicted temperature.  The departure from the analytic estimate observed in the $\tau_m=0.5$ case is a result of the inaccurate assumption that the source function is constant for $\tau\sim1$. 

\begin{figure}[ht]
   \centering
      \includegraphics[width=7cm] {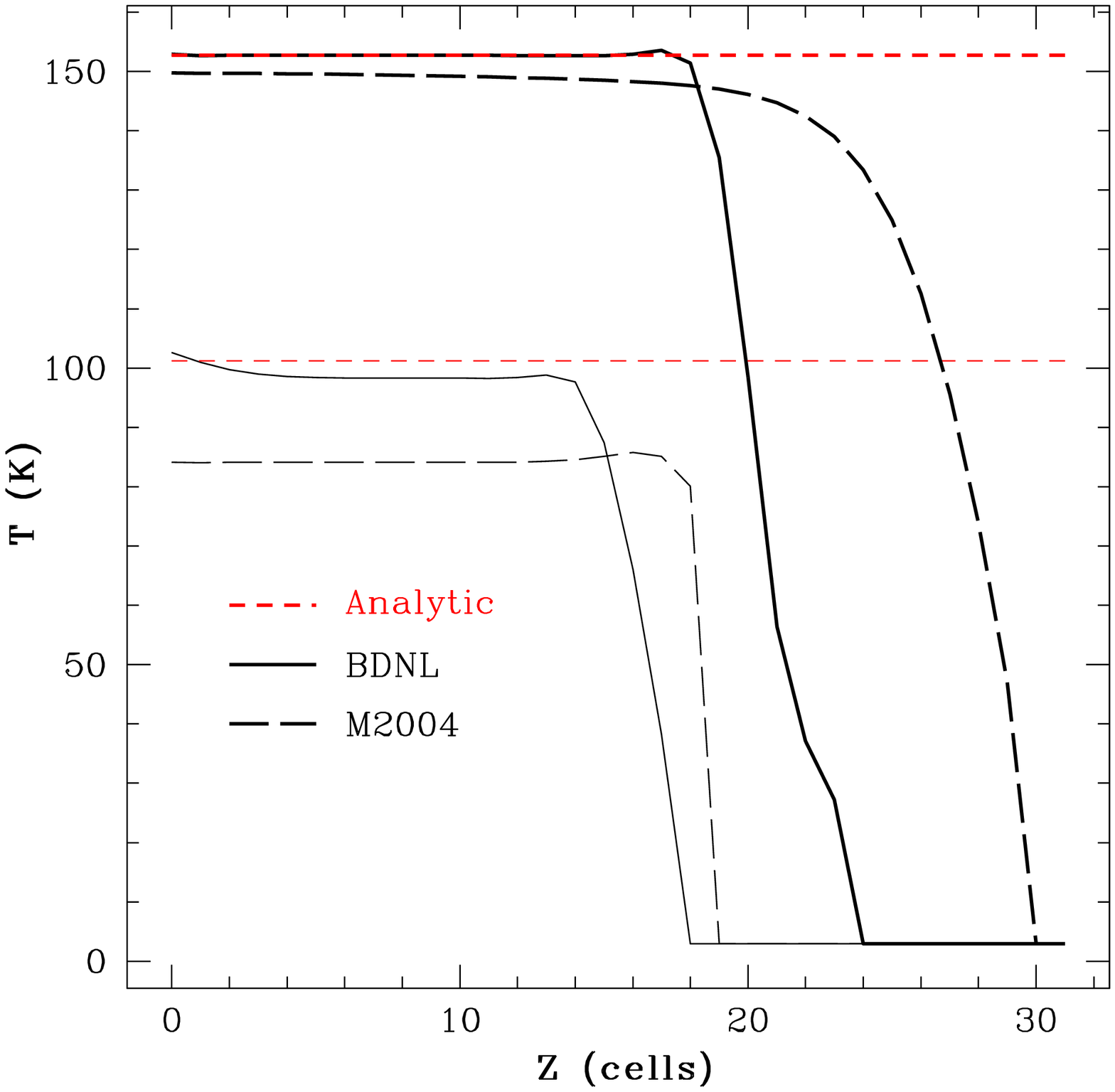}   \caption{Temperature curves for the low optical depth limit. The red, dashed curve indicates estimated isothermal temperatures from equation (18), the solid, dark curve indicates the results of the BDNL routine, and the dashed, dark curve indicates the results of the M2004 routine.  The curves at the lower temperature correspond to $\tau_m=0.5$, and the curves at the higher temperature correspond to $\tau_m=0.05$.  The lower optical depth corresponds to the higher temperature because cooling is less efficient.  The M2004 routine yields temperatures that are too cold, because it always uses the free-streaming approximation when $\tau<2/3$.  However, the M2004 routine converges to the correct solution as $\tau_m\rightarrow 0$.  The BDNL routine is consistent with the estimated temperature for the $\tau_m=0.05$ case, and it is roughly consistent with the $\tau_m=0.5$ case.  The inconsistency seen in the $\tau_m=0.5$ case is a result of the small $\tau_m$ assumption, which is used to derive equation (18).}
   \label{fig4}
\end{figure}

\subsection{Contraction Test}

In order to study how accurately the radiative algorithms work with the hydrodynamics routines and to study the effects of resolution, we allow the atmosphere to cool and follow the contraction.  If one assumes a constant opacity law and a large $\tau_m$, then the contraction becomes homologous, and a relationship between the midplane temperature and the cooling time is easily attainable.  Consider the cooling time
\begin{equation}t_{\rm cool}=\frac{U}{\sigma T_e^4}\sim \frac{p_m H \tau_m}{T_m^4},\end{equation}
where $H=\Sigma/\rho_m$ is the scale height of the atmosphere and $U$ is the internal energy per unit area.  If one assumes an ideal gas law, then one expects that
\begin{equation}t_{\rm cool} \sim \frac{1}{T_m^3}.\end{equation}
Finally, because $U\sim p_m H\sim T_m$,
\begin{equation}t_{\rm cool}\sim \frac{1}{U^3}.\end{equation}
For this test, we take the relaxed atmospheres shown in Figure 1 with $\tau_m=100$ and turn the heating term off.  The contraction is followed until the scheme breaks down.  Figure 5 indicates that both cases follow the expected contraction closely until the optically thick atmosphere is resolved by five  (BDNL) or six (M2004) cells. 

\begin{figure}[ht]
   \centering
   \includegraphics[width=7cm]{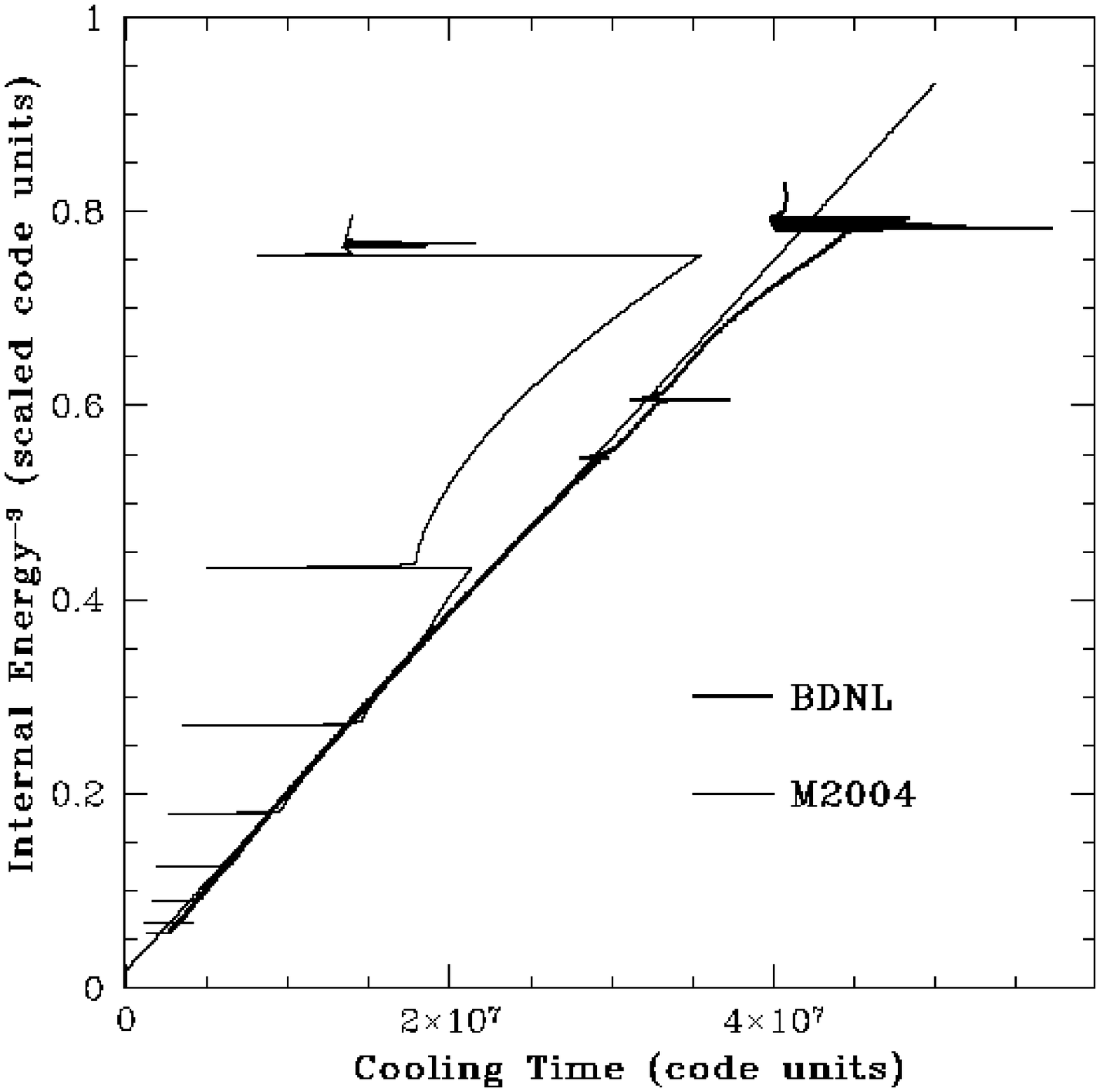}

   \caption{Contracting slabs as shown in the $t_{\rm cool}$ - $U^{-3}$ plane for the same cases as shown in Figure 1; all curves should be linear if the slabs contract as expected.  Both schemes break down when the high optical depth regime is contained within 5 cells ($U^{-3}\approx0.75$), but the M2004 routine (light curve) starts to deviate from the expected solution once the slab is resolved by 6 cells ($U^{-3}\approx0.43$).  The sudden decreases in cooling times for the M2004 routine are where the optically thin/thick boundary transitions into another cell.  The line is a by-eye fit to show the expected behavior. }
   \label{fig5}
\end{figure}

\subsection{Convection Test}

The last test we describe demonstrates whether the radiative scheme permits convection when it should occur.  Lin \& Paploizou (1980) and Ruden \& Pollack (1991) show that convection is expected in a disk-like atmosphere when the Roesseland mean optical depths are large and when $\beta > \beta_{\rm crit}$ for $\kappa\sim T^{\beta}$.  For a $\gamma=5/3$ gas, the vertical entropy gradient is driven to a negative value when the critical exponent $\beta_{\rm crit} \approx1.5$. So we present two cases for which almost identical atmospheres are allowed to relax to an equilibrium.  In one case, $\beta=1$, which should make the atmosphere convectively stable, and the other has $\beta=2$, which should make it unstable.  As shown in the top panels of Figure 6, we find that the BDNL routine produces convection when it should and does not when it should not.  Likewise, the bottom panels of Figure 6 demonstrate that the M2004 routine also permits or inhibits convection correctly.  However, M2004 does seed an artificial superadiabatic gradient at the boundary between the optically thin and thick regions.  Nevertheless, convection does not occur for $\beta=1$, and the superadiabatic gradients are an order of magnitude smaller for $\beta=1$ than for $\beta=2$ at that boundary. 

\begin{figure}[ht]
   \centering
   \includegraphics[width=7cm]{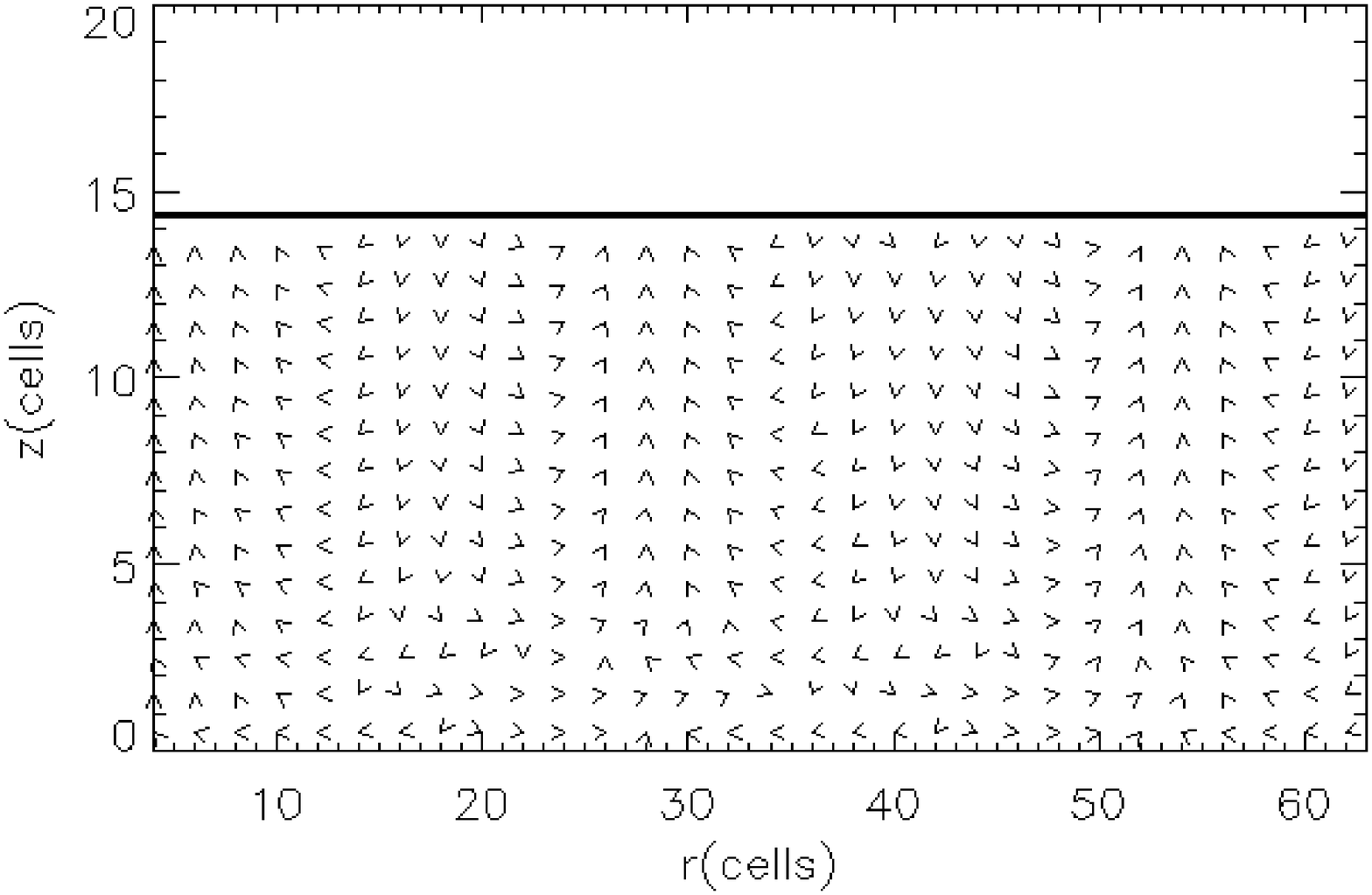}\includegraphics[width=7cm]{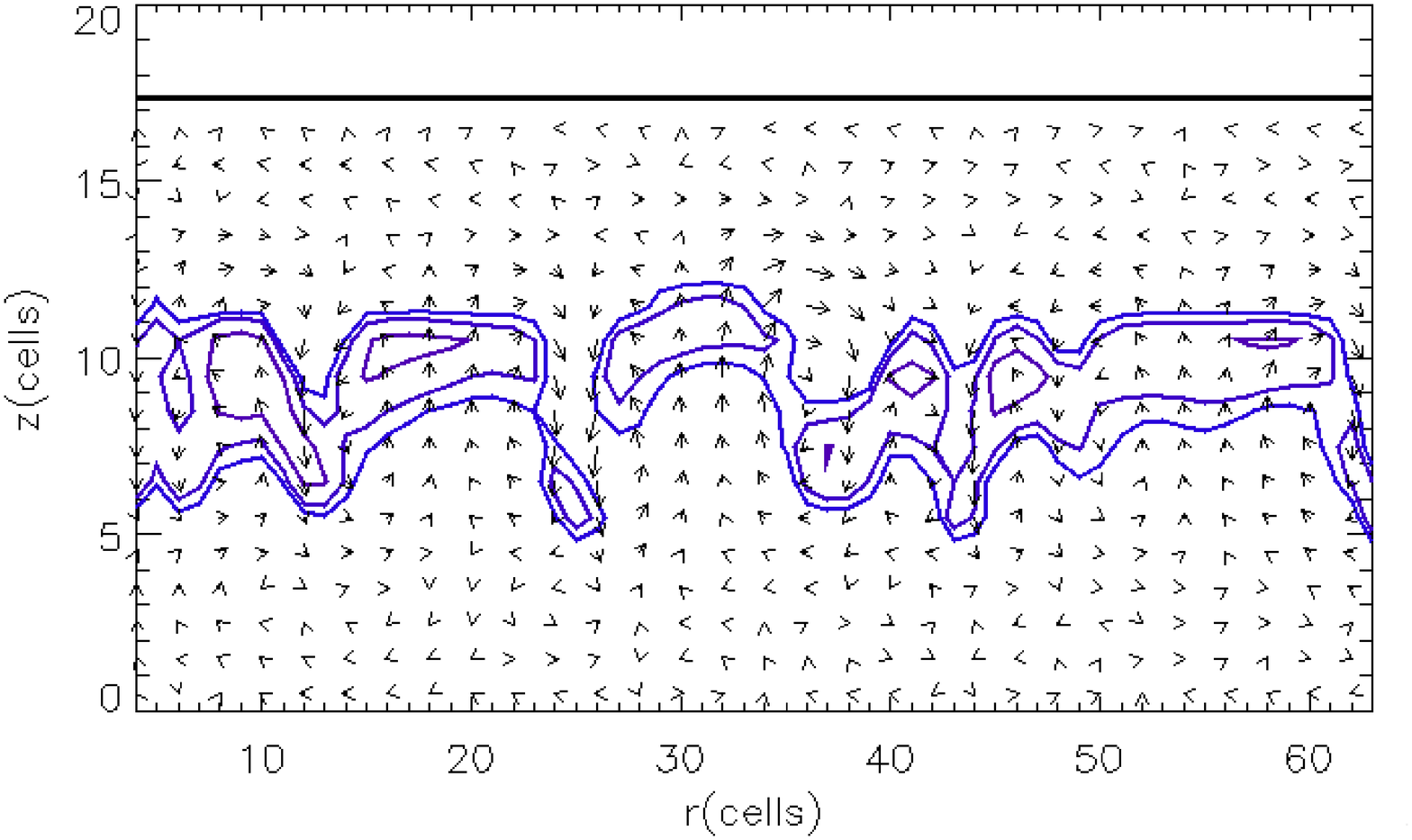} 
   \includegraphics[width=7cm]{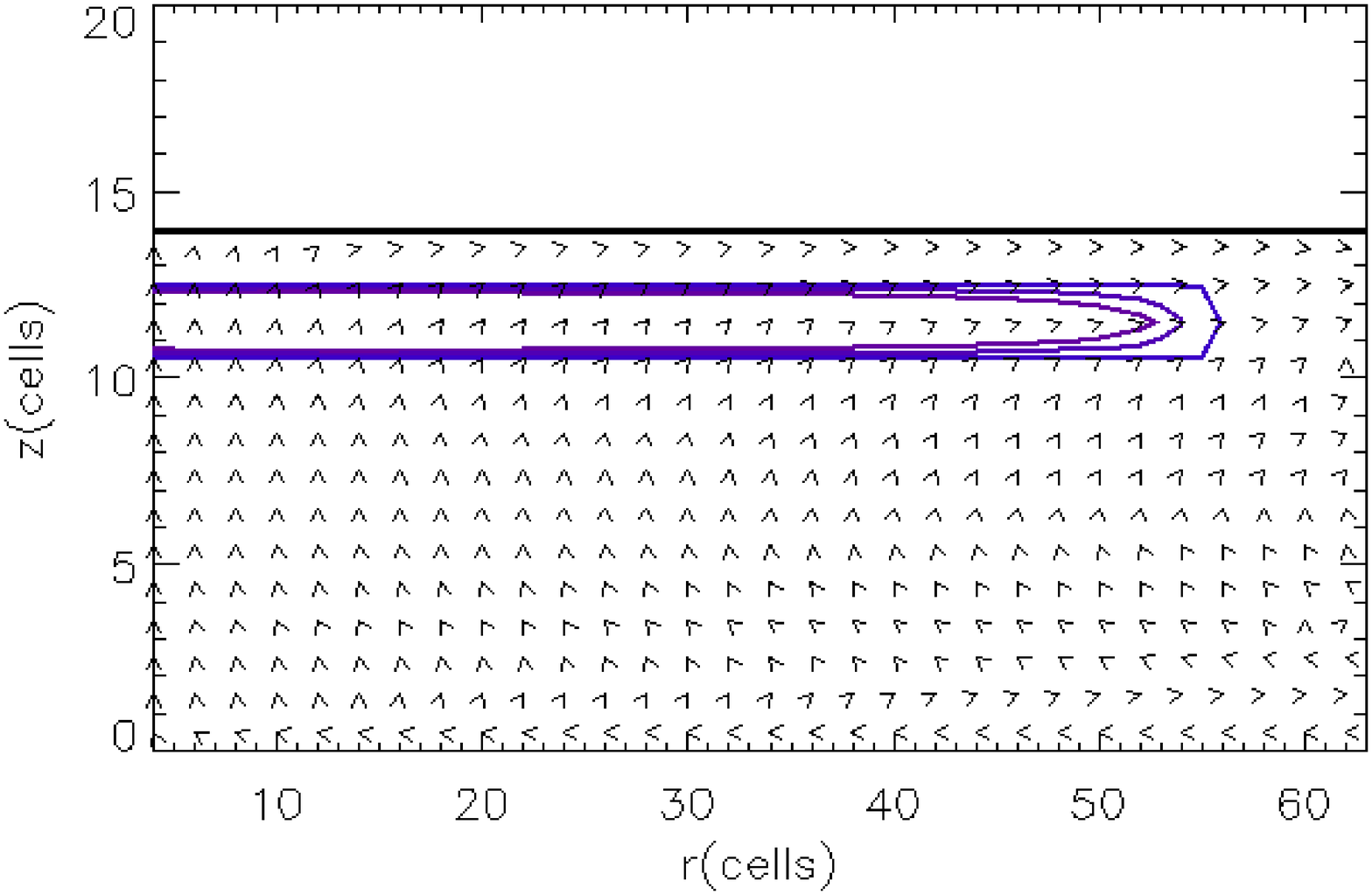}\includegraphics[width=7cm]{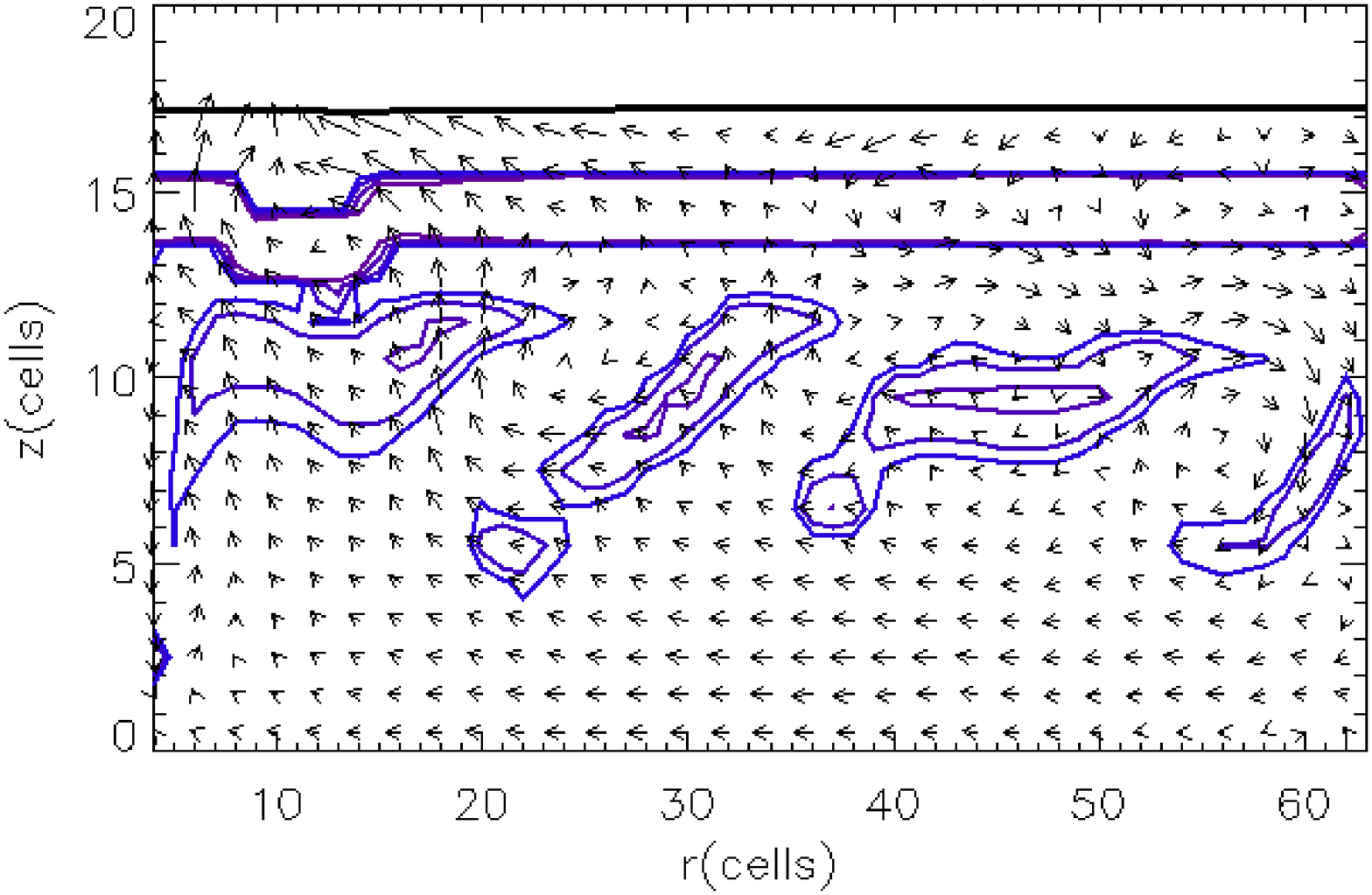} 

   \caption{Convection test.  The heavy black contour indicates the same density contour for each panel.  Arrows show relative velocities in the $r$ and $z$ directions.  The blue contours indicate superadiabatic regions.  The motions in the left panels are a few orders of magnitude smaller than the motions in the right panels. Top: BDNL. The left panel shows the case $\beta=1$; convection and superadiabatic gradients are absent, and the velocities represent low level noise.  The right panel shows the case $\beta=2$; convective cells and superadiabatic gradients are present. Bottom: Same as the top panel but for the M2004 scheme.  The superadiabatic regions near the top density contour are due to the artificial, sudden drop in temperature at the optically thin/thick interface.  The superadiabatic gradients in the left panel are about an order of magnitude smaller than those in the right panel.  }
   \label{fig6}
\end{figure}

\begin{figure}[ht]
   \centering
   \includegraphics[width=7cm]{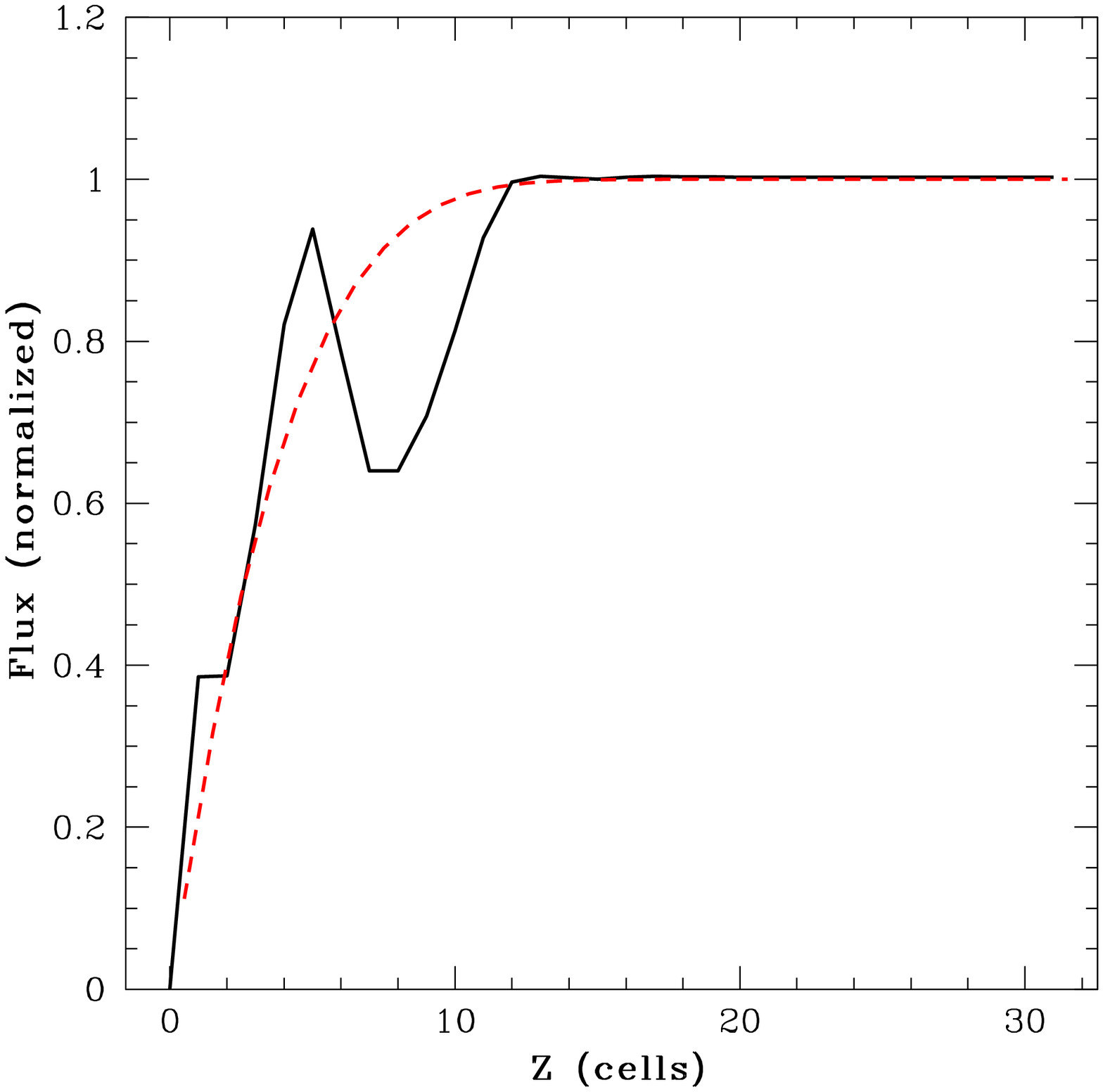}

   \caption{Average flux through the atmosphere with $\beta=2$ and with the BDNL scheme. Convection carries about 30\% of the flux at maximum, but almost all the flux is radiative in the photospheric region near cell 15.  }
   \label{fig7}
\end{figure}

Finally, we measure the flux that is carried by convection in the $\beta=2$ case for the BDNL scheme.  As indicated in Figure 7, convection carries about 30\% of the flux.  Figure 7 also acts as a reminder that the energy must ultimately be radiated away.  Paper I and Rafikov (2006) argue that because the convective flux is controlled by the radiative flux leaving the photosphere of the disk, convection should not be expected to lead to rapid cooling and fragmentation in protoplanetary disks, as claimed by Boss (2004) and Mayer et al.~(2007).

\section{SIMULATION}

In order to verify the results of Paper I and to test the dependence of disk evolution on the details of the treatment of radiative physics, we restarted the Paper I simulation at 6.6 orp\footnote{For these simulations, one outer rotation period (orp) corresponds to about 253 yr.}, but we evolved the simulation with the BDNL radiative algorithm. This time corresponds to just after the {\it burst phase}, which is characterized by a rapid and violent onset of global nonaxisymmetry in a gravitationally unstable disk.   We chose to restart the simulation at this time, because for this experiment, we are primarily interested in the behavior of the disk during the {\it asymptotic phase}, i.e., the phase of evolution during which heating roughly balances radiative cooling.  Moreover, beginning the simulation just after the burst is a compromise between lowering the computational cost of the simulation and obviating possible transients in the asymptotic phase that are caused by abruptly changing the cooling algorithm. 

The initial model of Paper I is the same as that used by Mej\'ia et al.~(2005), and we refer the reader to Mej\'ia et al.~for a full description.  The model is a 40 AU radius, 0.07 $M_{\odot}$ disk surrounding a $0.5$ $M_{\odot}$ star.  The initial disk has a $r^{-1/2}$ surface density profile and a $r^{-1}$ temperature profile, but these are significantly altered during the burst.   We use a constant ratio of specific heats $\gamma=5/3$ and the opacity tables of D'Alessio et al.~(2001), with a maximum grain size of 1 $\mu m$.  To control the numerical experiment, we use the same resolution that was used for the Paper I simulation, but the vertical direction is expanded to account for the more extended atmosphere the BDNL solution produces; $(r,~\phi ,~z)=(512,~128,~64)$ cells. In addition, the same erroneous mean molecular weight table, which typically yields $\mu=2.7$, was used (see Paper I for details).  

\subsection{Comparison between Disk Structures}

Although not extreme, qualitative differences between the simulations are discernible (Fig.~8). The Paper I simulation has more pronounced spiral structure than the BDNL simulation throughout most of the disk.  Also, the BDNL simulation is more extended in the vertical direction because the disk's atmosphere is hotter, as expected from the tests.    In order to quantify the structural differences, we compute the global Fourier amplitude spectrum (Imamura et al.~2000) for each simulation, where the sum over the amplitudes is a measure of the nonaxisymmetric structure in the disk and the spectrum is indicative of the dominate modes in the disk.  We compute the time-averaged Fourier component $\left<A_m\right>$ for $m$-arm structure by
\begin{equation}
A_m = \frac{\int\rho_m rdrdz}{\int \rho_0 rdrdz},
\end{equation}
\begin{figure}[ht]
   \centering
   \includegraphics[width=6.5in]{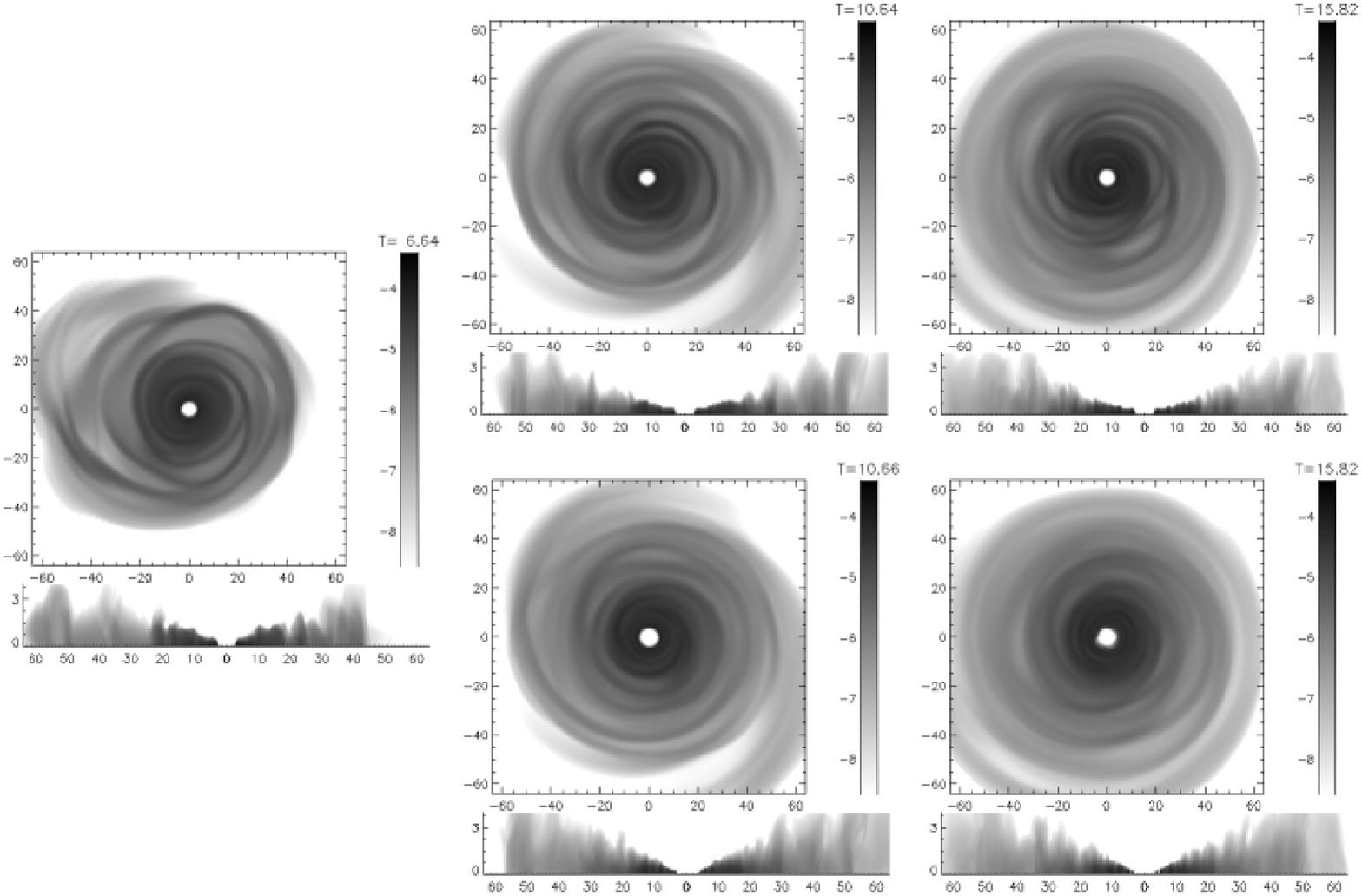}
   \caption{Midplane and meridional logarithmic density images for the Paper I simulation (top) and for the BDNL simulation (bottom).  The BDNL simulation begins from the Paper I simulation at 6.6 orp.  Although the differences are not extreme, they are readily noticeable. The BDNL simulation shows a more washed out structure overall.  It is also more extended vertically than the Paper I simulation.  Each axis is in AU, and the logarithmic grayscale indicates density in code units.  The number in the upper right of each image is the time in orp.}
   \label{fig8}
\end{figure}

\noindent where $\rho_0$ is the axisymmetic density component and $\rho_m$ is the total Fourier amplitude of the $\cos(m\phi)$ and $\sin(m\phi)$ density component.  The time-average is calculated by finding $A_m$ for a large number of snapshots over the last two orps. The summed global Fourier amplitude $\left<A_+\right>=\sum_{m=2}^{63}\left<A_m\right>=1.4$ for the Paper I disk, while $\left<A_+\right>=1.1$ for the BDNL disk, which indicates that the Paper I disk is more nonaxisymmetric.  We exclude $m=1$ from the summation because we keep the star fixed.  The difference between the sums is also depicted by the Fourier spectrum (Fig.~9).  The Paper I disk has larger amplitudes everywhere except for $m=2$, which is consistent with the qualitative differences portrayed in Figure 8.  Including long-range transport has resulted in weaker GIs. This is also consistent with the results of Cai et al.~(2007, in preparation), who find that when mild envelope irradiation is included, GIs are weakened as a whole, but the $m=2$ mode remains strong.  As described in Paper I,  a curve with the  functional form $A_m\sim\left(m^2+m_0^2\right)^{-n}$ can be fit to the data, where $n\approx1.6$.  The BDNL disk is also consistent with this functional form, and both disks are roughly consistent with by-eye fits of $n\approx 1.5$.  The similar slopes at large $m$ may be indicative of gravitoturbulence Gammie (2001).  However, whether this slope is caused by nonlinear mode coupling or a turbulent cascade is a topic for future discussion.

\begin{figure}[ht]
   \centering
   \includegraphics[width=6.5in]{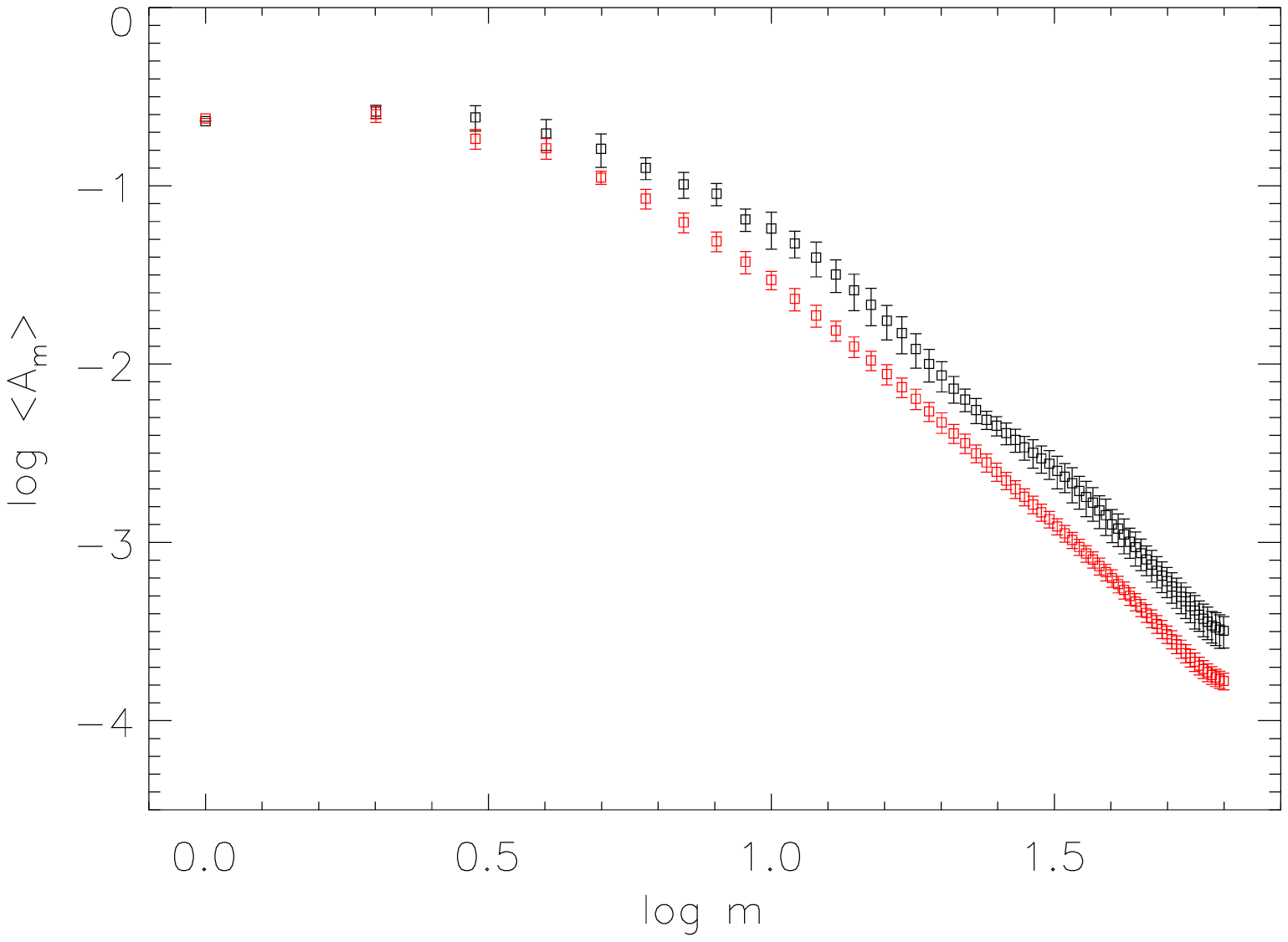}
   \caption{Fourier amplitude spectrum for the Paper I (black) disk and the BDNL (red [gray]) disk, time averaged over the last two orp of each simulation.  The bars represent typical fluctuations over the two-orp period.  The Paper I has larger amplitudes everywhere except for $m=2$, which is consistent with the Paper I disk being more nonaxisymmetric.  Both spectra can be fit with a functional form of $A_m\sim(m^2+m_0^2)^{-n}$, where $n\approx 1.5$.  This behavior at large $m$ may be indicative of gravitoturbulence (Gammie 2001).}
   \label{fig9}
\end{figure}

The surface density profiles (not shown) for the end of each simulation are comparable and follow a Gaussian profile.   A ring forms in both simulations near 7 AU. However, this ring appears to be caused by poor vertical resolution.  For $r\ge10$ AU, the disk is resolved well vertically, and the radial mass concentrations that form due to multiple, tightly wrapped spiral arms are reliable. These mass concentrations can work to concentrate solids (Haghighipour \& Boss 2003; Rice et al.~2004), which may act to accelerate the core accretion plus gas capture mechanism of gas giant planet formation(Durisen et al.~2005). 

\subsection{Comparison between Disk Energetics}

Figure 10 shows the evolution of the internal energy for each disk.  There is a precipitous drop in energy over about an orp after switching the radiative transport schemes. However, the schemes roughly follow each other after the drop, with the BDNL profile having a more shallow slope than the Paper I profile.  
The effective temperature profiles (Fig.~11) are also similar.  Each profile can be fit by an exponential, and as discussed in Paper I, we believe that the deviation from the observationally expected $r^{-0.5}$ profile probably results from our exclusion of stellar irradiation in these simulations.  

\begin{figure}[ht]
   \centering
   \includegraphics[width=6.5in]{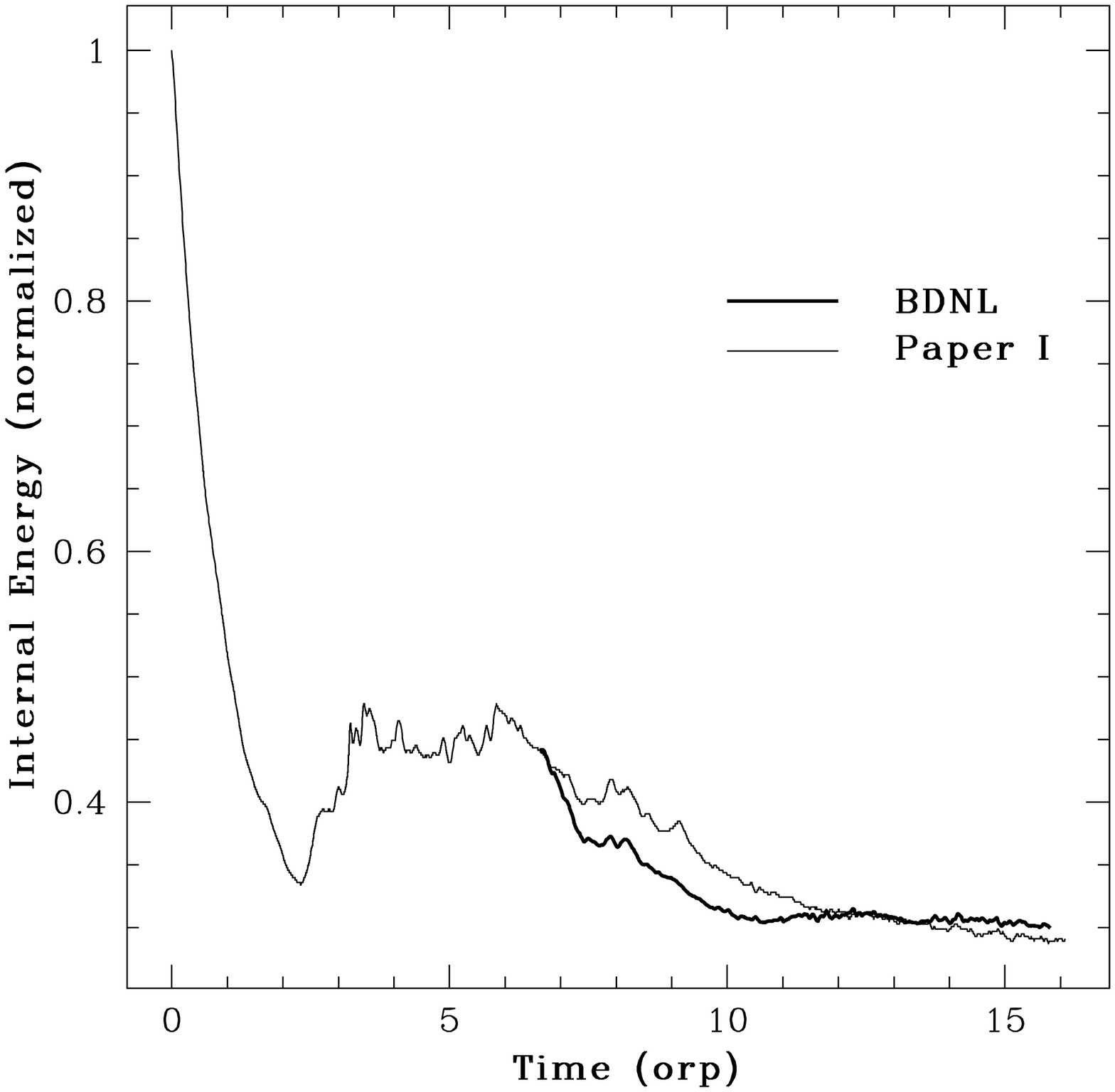}
   \caption{Internal energy normalized to the initial value for BDNL (heavy curve) and Paper I (light curve).  The precipitous drop is a result of suddenly switching radiative schemes.  Between about 7 and 10.5 orp the curves approximately track each other.}
   \label{fig10}
\end{figure}
 
\begin{figure}[ht]
   \centering
   \includegraphics[width=6.5in]{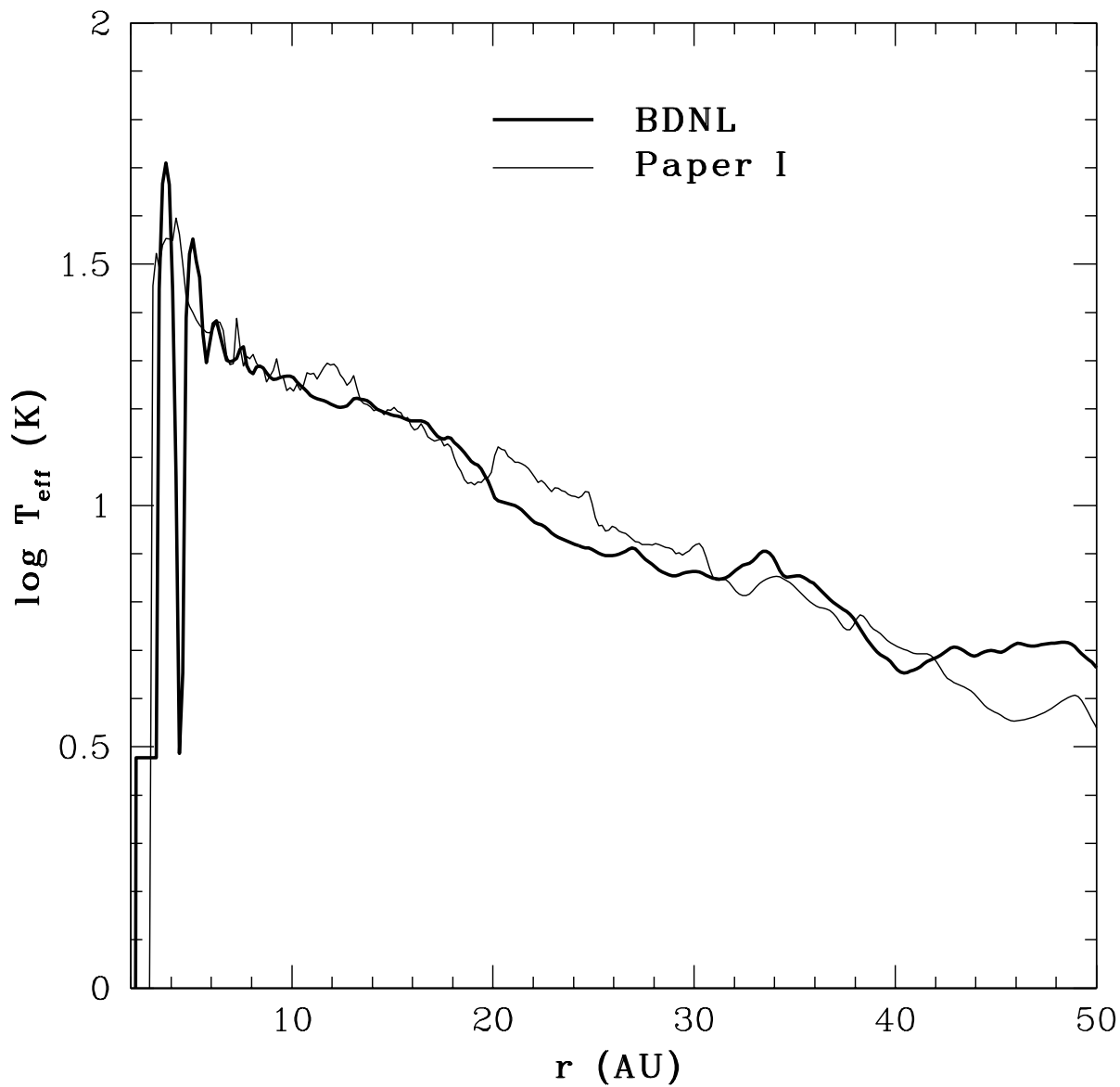}
   \caption{Effective temperature profiles for BDNL (heavy curve, time-averaged over about the last 5 orp) and Paper I (light curve, time-averaged over the last 6 orp).  Both follow an exponential profile, and are reasonably consistent.  Their departure from the observed $r^{-1/2}$ effective temperature profiles is likely due to our exclusion of stellar irradiation.}
   \label{fig11}
\end{figure} 

\begin{figure}[ht]
   \centering
   \includegraphics[width=6.5in]{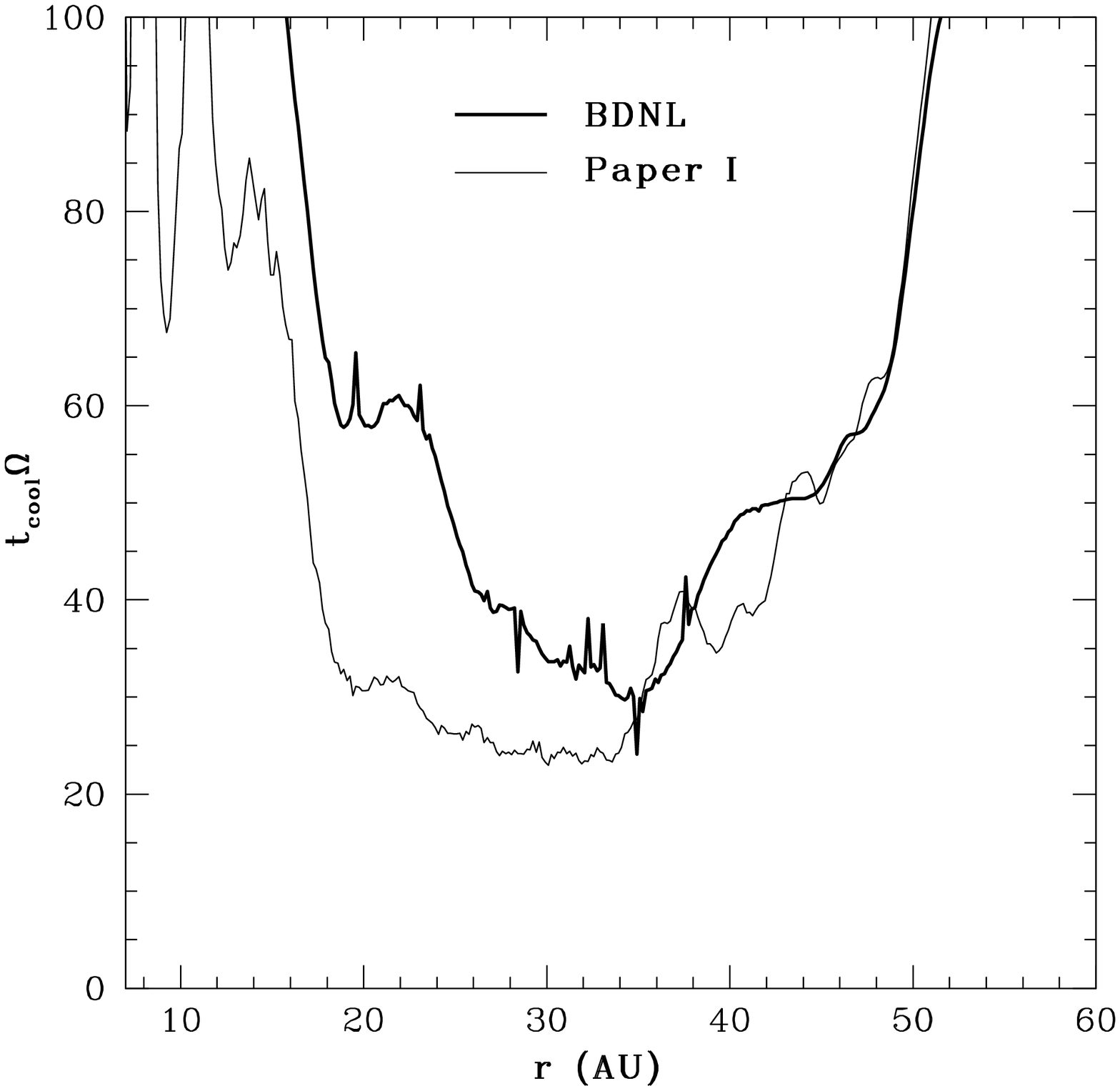}
   \caption{Cooling time curves scaled by the local angular speed for BDNL (heavy curve) and Paper I (light curve) for the time-averaged periods over about the last 5 and 6 orp, respectively.  Both curves are relatively consistent for $r\gtrsim 35$, but they depart for inner radii.  This is likely due to a combination of the free-streaming approximation, employed by the Paper I simulation, in regions where long-range coupling matters and the different opacities used by the two routines.}
   \label{fig12}
\end{figure} 

For each simulation, we calculate the time-averaged cooling time 
$
t_{cool} = \int \epsilon dV/\int \nabla \cdot F dV,
$
for each annulus on the grid, where $\epsilon$ is the internal energy density of the gas and $\nabla \cdot F $ is the total radiative cooling. The temporal average is taken to be about the last six orp of evolution for Paper I and about the last 5 orp for BDNL.  In Figure 12, we compare $t_{cool}\Omega$ curves for each disk, where $\Omega$ is the angular speed of the gas.  The cooling time is much longer for $r \lesssim 35$ AU in the BDNL disk than it is for the Paper I disk.  However, the curves converge outside that radius.   The longer cooling times are consistent with the washed out structure in the BDNL simulation.  For both disks, the cooling times are well above the fragmentation criterion $t_{cool}\Omega \lesssim 6 $ for a $\gamma=5/3$ gas (Gammie 2001; Rice et al.~2005), so we expect neither disk to fragment.  Regardless, we ran the BDNL simulation at 512 azimuthal divisions between 10 and 11 orp to test for fragmentation and found no signs of fragmentation, as one expects from the long cooling times.

\subsection{Comparison between Angular Momentum Transport}

In this section, we compare the angular momentum transport in each disk by analyzing the gravitational torque on the inner disk due to the outer disk and by measuring the effective Shakura \& Sunyaev (1973) $\alpha$.  As discussed in Paper I, the gravitational torque is calculated by 
\begin{equation} C = -\int \rho {\bf x} \times {\bf \nabla}\Phi~dV, \end{equation}
where $\Phi$ is the gravitational potential, ${\bf x}$ is the position vector, and the integral is over the entire volume.  For our analysis, we are concerned with the vertical component $C_z$.  The time-averaged torque, averaged over the last six orp for Paper I and about the last 5 orp for BDNL, is shown for each simulation in Figure 12.  The solid curves represent the torque profiles, with the heavy curve indicating the BDNL disk.  The dashed curves show the mass flux for each disk with arbitrary but consistent scaling; the peak mass flux $\dot{M}=$ few$\times10^{-7}~M_{\odot}~\rm yr^{-1}$ for each disk.  The torques are of the same magnitude, but the torque profiles are noticeably different.  Based on the $\left<A_m\right>$ plots and the visual differences in disk structures, the Paper I disk has a more complex morphology and stronger modes. The multiple peaks in the Paper I torque profile are another indication of this complex morphology and competing global, dominate modes.  In contrast, the BDNL disk has a more washed out structure, and so its torque profile has only a single maximum.

Both disks have complicated mass flux profiles.  The principal inflow/outflow boundary in the Paper I simulation is at about $r\approx 26$ AU.  The BDNL disk, by contrast, has two main inflow/outflow boundaries.  The $r\approx15$ AU  boundary corresponds to the peak torque in the BDNL disk, and we refer to this as the principal inflow/outflow boundary because the mass fluxes are the highest near it.  Roughly, the peak in each torque profile aligns with the principal inflow/outflow boundary.  The agreement is imprecise, because the mass flux average is based on differencing mass cylinders at different times, which yields a time average based on the second-order mass flux integration.  The torques are derived in post-analysis calculations, and so the temporal sampling is much sparser.  Moreover, the mass fluxes are highly variable with time, and averages over slightly different time periods can result in different mass flux profiles.  However, major inflow/outflow transitions are usually near torque profile maxima. The mass fluxes for $r\gtrsim 40$ are complicated by pulsations that begin just before the disk bursts and continue throughout the evolution.  

The gravitational torque can also be used to derive an effective $\alpha$.
We use this torque to calculate the vertically integrated gravitational stress $T$, where $T=C_z/2\pi r^2$ (Paper I).  This stress can be used to calculate an effective $\alpha$ (Gammie 2001)
\begin{equation} \alpha =   \Bigl\lvert\frac{d\ln \Omega}{d\ln r}\Bigl\lvert^{-1}\frac{T}{\left< c^2 \Sigma \right>},\end{equation}
where the brackets indicate an azimuthally averaged quantity, $c$ is the midplane sound speed, and $\Sigma$ is the surface density.  We use the adiabatic sound speed for consistency with Paper I even though the isothermal sound speed may be more appropriate (e.g., Balbus \& Papaloizou 1999; Gammie 2001).  The comparison between the effective $\alpha$ for BDNL and Paper I is shown in Figure 13.  The profiles are of similar magnitude everywhere, but with BDNL being significantly lower between about 20 and 36 AU. 

We also show in Figure 13 the $\alpha$ one would expect for an $\alpha$ disk (see Gammie 2001)
\begin{equation}
\alpha =  \left(\Bigl\lvert\frac{d\ln\Omega}{d\ln r}\Bigl\lvert^2 \gamma'\left[\gamma'-1\right]t_{cool}\Omega\right)^{-1},
\end{equation}
where $\gamma'$ is the two-dimensional adiabatic index.  For $\gamma=5/3$, $\gamma'\approx 1.8$ in a strongly self-gravitating disk and 1.5 in a non-self-gravitating disk (Gammie 2001).  We use the $t_{cool}\Omega$ profiles (Fig.~12)  in equation (27) to plot the anticipated $\alpha$ for a local model in the non-self-gravitating limit.  For most radii, both disks are roughly consistent with this $\alpha$ prescription, and $\alpha$ is roughly constant between 20 and 35 AU.   The main difference between the two simulations is the lower $\alpha$ in the BDNL simulation, which is consistent with the longer cooling times.  These different cooling times are probably due to a combination of the different degrees of cell-to-cell coupling and the use of different opacities. 

Although the overall evolutions are in rough agreement, they demonstrate sensitivity to the details of radiative transfer.  By including the long-range effects of radiative transfer and by using different opacities for different optical depth regimes, the BDNL disk shows less structure, is more flared, and has a lower effective $\alpha$ that deviates slightly more from equation (27).  These differences demonstrate the need for a radiative algorithm that includes the long-range effects of radiative transfer in all three-dimensions, which will be missed by diffusion approximations and is missed in our schemes in the $r$ and $\phi$ directions.  To illustrate the importance of radiative transport in all three directions, we show in Figure 15 the effective $\alpha$ profile for a disk that was evolved with the BDNL radiative routine, but with the radial and azimuthal diffusive radiative transport turned off.  According to this plot, without any $r$ and $\phi$ transport, we would underestimate the cooling times in the optically thick regime and surmise that the disk deviates strongly from the predicted effective $\alpha$.

\begin{figure}[ht]
   \centering
   \includegraphics[width=6.5in]{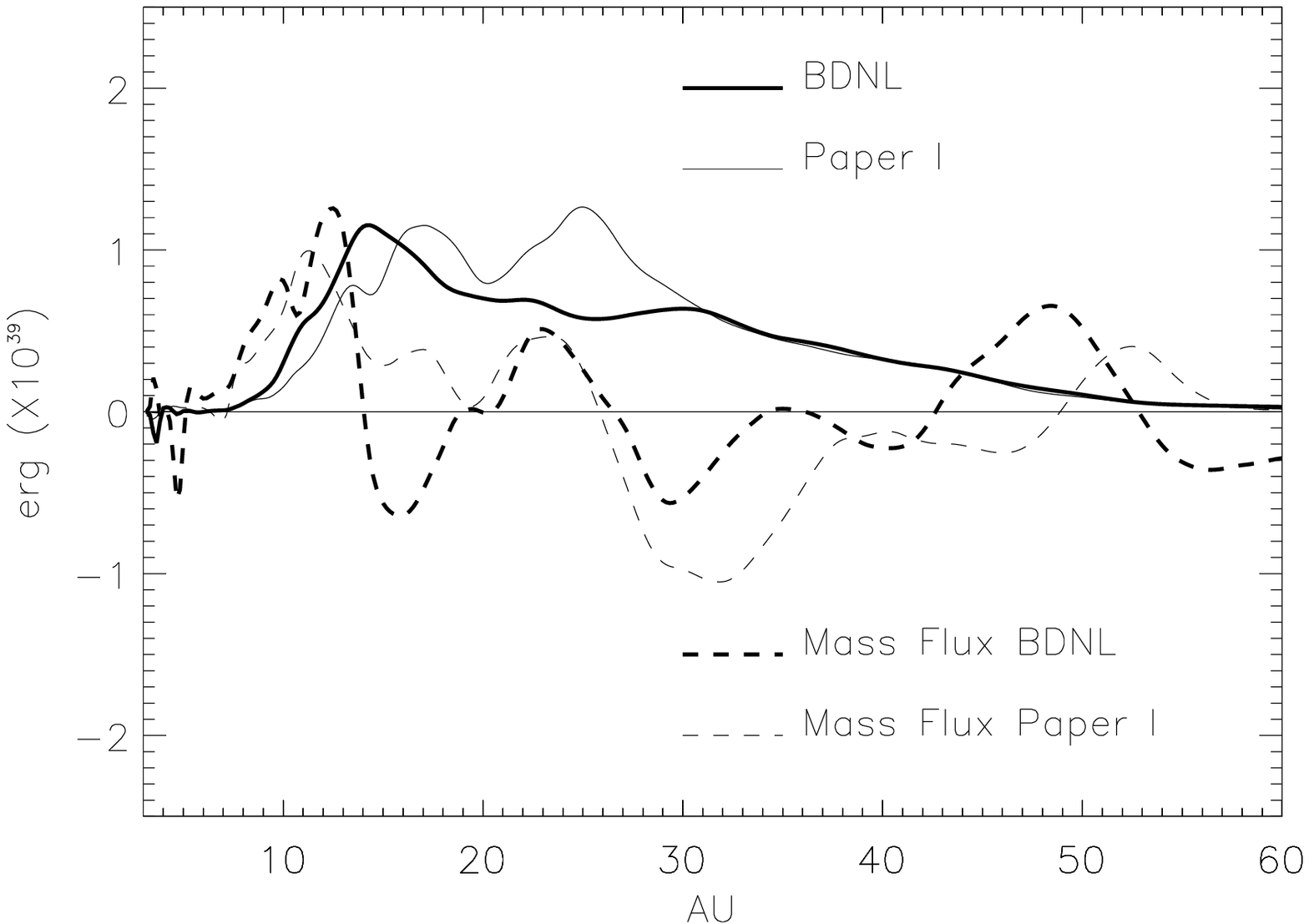}
   \caption{Gravitational torque (solid curves) and mass flux (dashed curves) profiles for BDNL (heavy curves), time-averaged over the last 5 orp, and Paper 1 (light curves), time-averaged over the last 6 orp.  BDNL's torque profile shows one strong peak and several minor peaks, while the Paper I torque profile has two very strong peaks.   The mass fluxes for each disk (arbitrary but consistent scaling) are consistent in magnitude.  }
   \label{fig13}
\end{figure}
 
\begin{figure}[ht]
   \centering
   \includegraphics[width=6.5in]{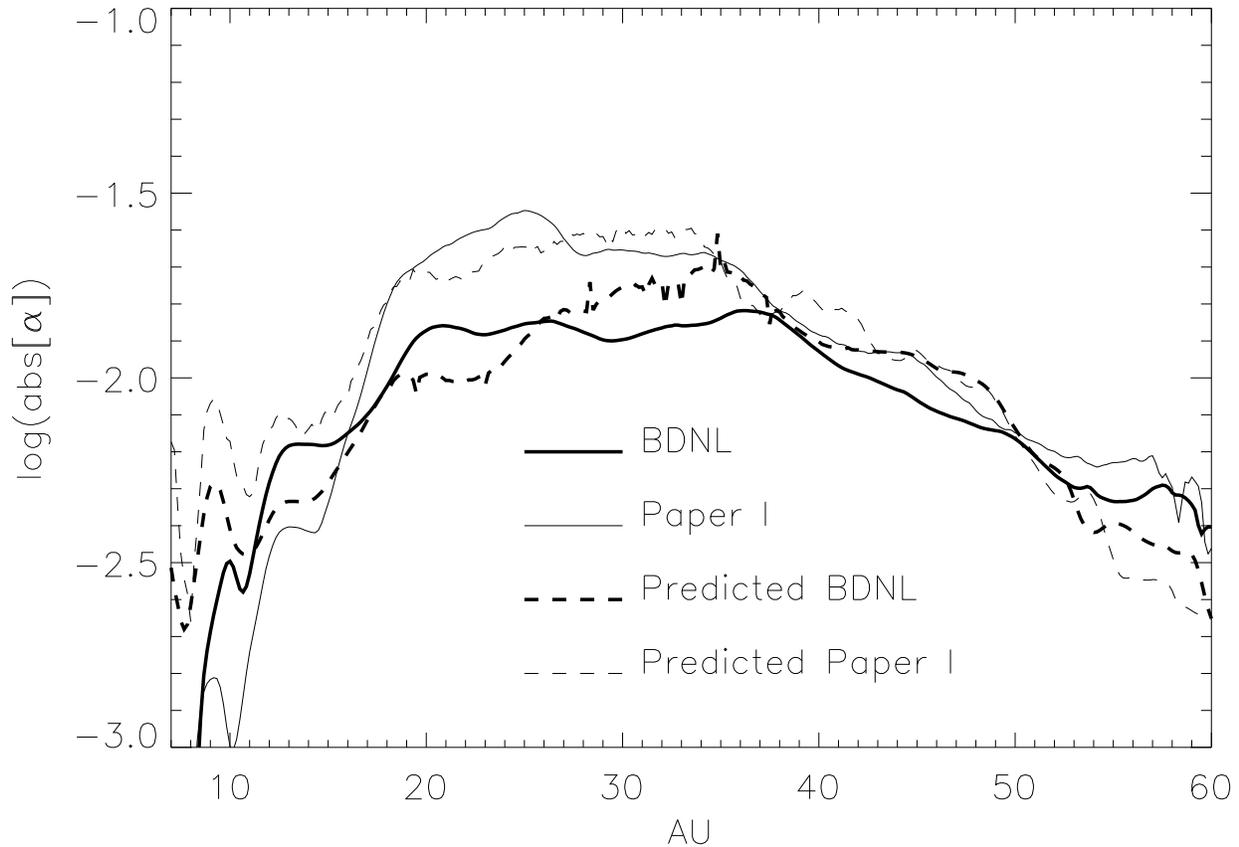}
   \caption{Effective $\alpha$ profiles for BDNL (heavy curves) and Paper I (light curves). The solid curves indicate the effective $\alpha$ derived from the torque profiles, and the dashed curves indicate the predicted $\alpha$ based on an $\alpha$ disk prescription, for which the predicted $\alpha$ is derived from the $t_{cool} \Omega$ profiles (Gammie 2001) in Figure 12 with the assumption of negligible self-gravity (see text).  Both disks roughly follow the predicted $\alpha$ over a larger range of radii. }
   \label{fig14}
\end{figure} 

\begin{figure}[ht]
   \centering
   \includegraphics[width=6.5in]{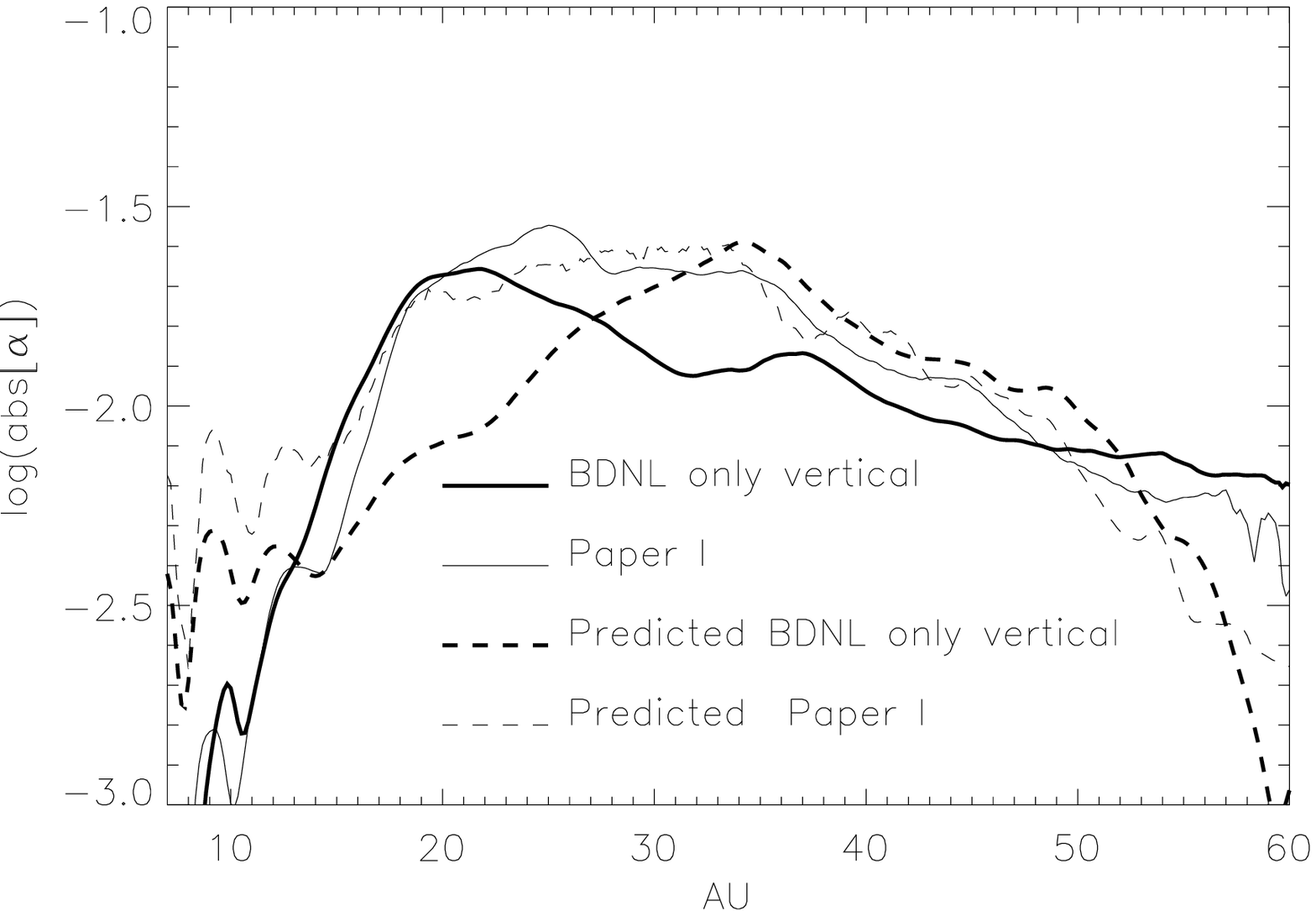}
   \caption{Same as Figure 14, but the BDNL simulation has only vertical radiative transport. }
   \label{fig15}
\end{figure} 

\section{CONVECTION}

Boss (2001) and Mayer et al.~(2007) report very short cooling times in their disks, and they attribute these fast cooling times to convection (Boss 2004).  This is contrary, however, to what Ravikov (2006) predicts analytically and to what is reported in Paper I, where convection does not lead to cooling times short enough for fragmentation.  Paper I and Ravikov (2006) argue that convection should not lead to fast cooling times in protoplanetary disks because the energy transported by convection must ultimately be radiated away near the photosphere of the disk.  The BDNL disk is stable to convection in the high-optical depth regions, and the superadiabatic regions of the Paper I disk\footnote{Although Paper I found some superadiabatic regions in their disk during the asymptotic phase, convection was not observed because the spiral waves dominated the dynamics and the superadiabatic regions were in regions where $\tau\sim1$.} are mostly related to the artificial temperature drop at the photosphere that is characteristic of the M2004 scheme.  In order to make relevant comments regarding convection in this paper, we briefly discuss a disk model with parameters tuned to induce convection.  

The model that we explore is of a 10 AU in radius, optically thick,  approximately 0.1 M$_{\odot}$ disk around 1 M$_{\odot}$ star.  We set $\gamma=1.4$ and the opacity law $\kappa = (T/150 K)^3$.  This opacity power law, along with the chosen $\gamma$, ensure that the disk should be convective (see \S 4.3) as long as no other dynamics are present.  The model is moderately stable to GIs, with $Q\approx 1.8$ for most radii.  
Figure 16 shows that convection appears to be very active in this model. There are large pockets of negative entropy gradients at mid-disk altitudes.  Crude measurements of energy transport by vertical 
gas motions $F_c=\rho c_p v_z \Delta T$ indicate that more than half of the energy can be carried by these motions in the low to mid-disk altitudes.  Here, $c_p$ is the specific heat at constant pressure, $v_z$ is the vertical gas velocity, and $\Delta T$ is the deviation of the temperature at the cell of interest from the mean temperature for a given annulus and height.  Even though a large fraction of the energy in the low to middle regions of the disk can be transported by vertical gas motion, $t_{cool}\Omega\approx1000$ near 5 AU!  The energy must ultimately be radiated away near the photosphere, and so the cooling times remain long because the cooling time at the photosphere regulates convection.  This result is consistent with Ravikov's (2006) analytic predictions, with our numerical tests, our numerical simulations, and the numerical simulations of Nelson et al.~(2000) and Nelson (2000). We also point out that Nelson et al.~(2000) assume a vertically isentropic density profile when calculating the cooling times for their 2D SPH calculations, which is similar to assuming efficient convection.  We suspect that Boss (2004) and Mayer et al.~(2007) see fast cooling due to convection because of the different treatment of the photospheric layers in their simulations.  As discussed by Nelson (2006) and in this paper, proper treatment of radiative physics, especially near the photosphere, is crucial for estimating proper cooling times.  In fact, in order to lower the cooling times for the calculation in this section to those expected to lead to fragmentation, $t_{cool}\Omega\lesssim12$ for a $\gamma=7/5$ gas  (Rice et al.~2005), the effective temperature would need to be approximately equal to the midplane temperature, which is about three times the actual effective temperature at $r=5$ AU.  

\begin{figure}[ht]
   \centering
   \includegraphics[width=6.5in]{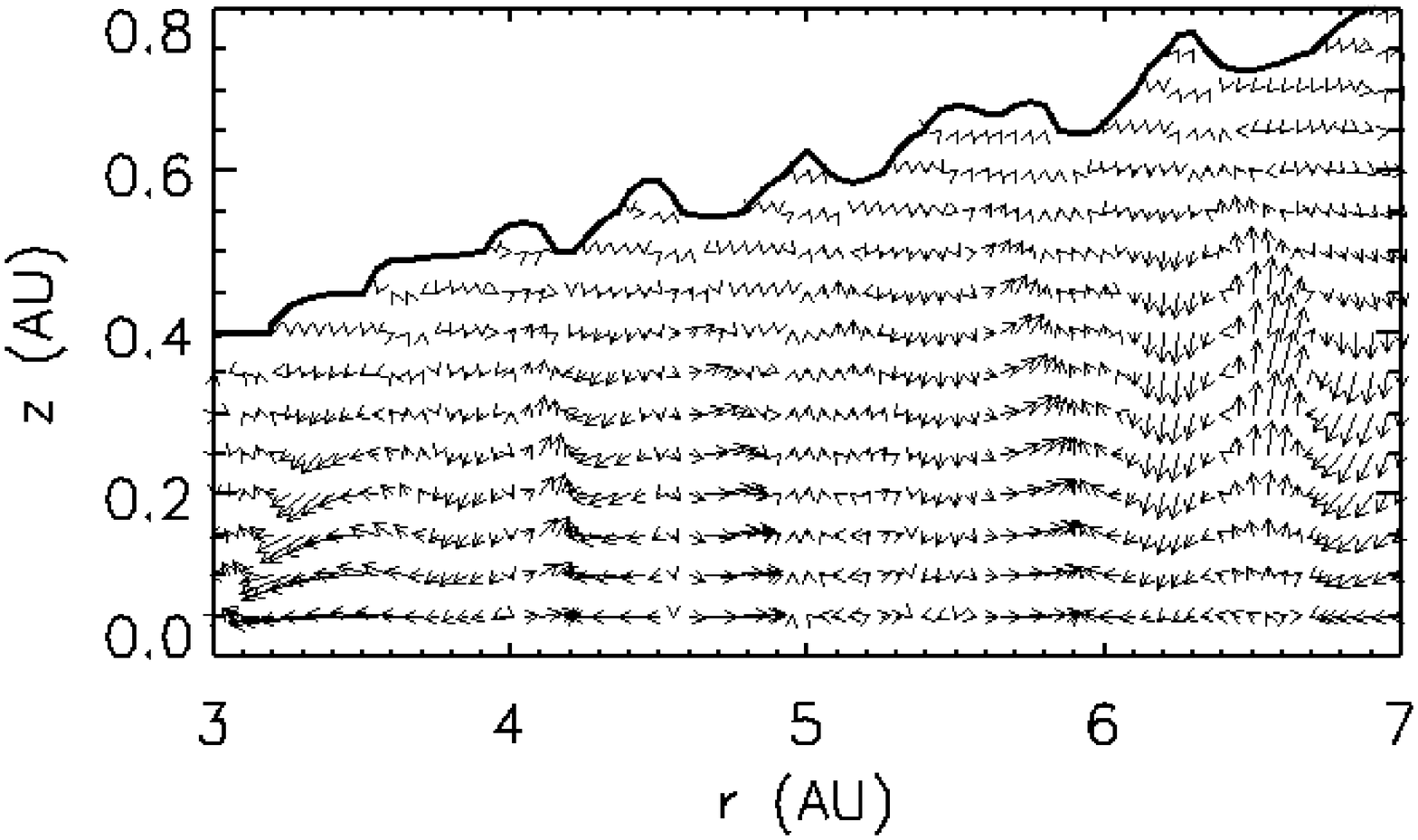}
   \caption{Convection-like motions in an optically thick protoplanetary disk model (see text in \S 6).  The heavy curve roughly indicates the disk's photosphere, and the arrows, which are scaled to each axis and to the midplane density for each column, indicate the momentum density.  Typical Mach numbers for the gas range between a few hundredths to a few tenths. Convective-like eddies are present throughout most of the disk, and in the 5 AU region, vertical motions can carry most of the flux to upper disk altitudes.  However, cooling times remain long in this disk, because ultimately, the energy must be radiated away.}
   \label{fig16}
\end{figure}

 \section{SUMMARY}
 
To help evaluate the accuracy of radiative transport schemes in protoplanetary disk simulations, this paper presented a test suite that assesses an algorithm's accuracy (1) in matching analytic temperature and vertical flux profiles for a simple geometry with a dissipation-like heating source, (2) in following the expected contraction sequence for a slab undergoing quasi-static gravitational contraction, and (3) in permitting and inhibiting convection under the appropriate conditions.  We used this suite to test the M2004 and the BDNL radiative transfer algorithms and presented the results.  We recognize that even if an algorithm passes all of our tests, it does not guarantee that the algorithm will always be accurate.  However, if an algorithm cannot pass our tests, simulations using that algorithm are probably untrustworthy.

The BDNL scheme is an improvement over the M2004 scheme because it includes cell-to-cell coupling in the vertical direction.  This difference leads to the correct temperature structure for the disk, according to our tests, in both the optically thick and thin regimes, while the temperature structure for the M2004 scheme is too cool in the optically thin regime.   

To investigate possible consequences for disk evolution when employing the BDNL radiative scheme and to verify the results of the Paper I simulation, we evolved the Paper I disk with the BDNL algorithm.  Even though the two schemes have many similarities, the BDNL simulation shows less structure overall.  This indicates that radiative schemes that employ pure flux-limited diffusion, which excludes long-distance coupling, will likely behave differently from ray-based schemes.  This also indicates that there is much room for improving the BDNL scheme. A fully three-dimensional ray method (e.g., Heinemann et al.~2006) would be more realistic and should wash out the GIs even more than in the BDNL simulation. The mass transport in the disk is roughly consistent with the effective $\alpha$ predicted by Gammie (2001).   Even so, we note that the picture of mass slowly diffusing through the disk is misleading. As indicated in Figure 9 and in \S 5.3, mass transport is dominated by global modes, with large fluctuations in the mass fluxes at any given radius.  

Overall, we verify the basic results of Paper I.  Cooling times are long, and the disks are stable against fragmentation.  GIs are efficient at transporting angular momentum, with effective $\alpha\sim10^{-2}$.  In addition, our simulations agree with analytic predications that convection should not lead to rapid cooling and fragmentation.  
 
Why do researchers disagree on key properties of disk evolution? Boley et al.~(2007) suggest that the treatment for the internal energy of H$_2$ may be a contributing factor.  The results of the comparisons presented here strongly suggest that radiative physics is another likely cause. In fact, the sensitivity of disk evolution to radiative transfer details herein reported indicates that the radiative cooling algorithms may be the {\it primus inter pare} of causes.

\acknowledgments{We would like to thank A.~Boss, K.~Cai, C.~Gammie, L.~Mayer, S.~Michael, M.~Pickett, T.~Quinn, D.~Stamatellos, and A.~Whitworth for useful discussions and comments during the preparation of this manuscript.  A.C.B.'s contribution was supported by a NASA Graduate Student Researchers Program fellowship.  Contributions by R.H.D., \AA.N., and J.L.~ were supported by NASA grant NNG05GN11G, by the Danish Agency for Science, Technology and Innovation and the Danish Center for Scientific Computing, and by the National Science Foundation through grant AST-0452975 (astronomy REU program to Indiana University), respectively. This work was also supported in part by systems obtained by Indiana University by Shared University Research grants through IBM, Inc.~to Indiana University.
 
\clearpage


\bibliographystyle{apj}

\end{document}